\documentclass{aims}
\usepackage{amsmath}
  \usepackage{paralist}
  \usepackage{graphics} 
  \usepackage{epsfig} 
\usepackage{graphicx}  \usepackage{epstopdf}
 \usepackage[colorlinks=true]{hyperref}
\hypersetup{urlcolor=blue, citecolor=red}

\def\currentvolume{X}
 \def\currentissue{X}
  \def\currentyear{200X}
   \def\currentmonth{XX}
    \def\ppages{X--XX}
     \def\DOI{10.3934/xx.xx.xx.xx}

\theoremstyle{definition}

\usepackage[pagewise]{lineno}

\usepackage{amssymb,amsthm,MnSymbol}
\usepackage{color}

\usepackage{algorithm}
\usepackage{algorithmic}

\usepackage{subfig}

\newcommand{\dd}[2]{\frac{\diff#1}{\diff#2}}

\DeclareMathOperator{\diff}{d}

\usepackage{amscd}
\usepackage{mathrsfs}

\def\dd{{\color{red}\mathsf d}}

\usepackage{enumitem}

\definecolor{darkgreen}{rgb}{0.0,0.5,0.0}               
\definecolor{lightblue}{rgb}{0.3296, 0.6648, 0.8644}    
\definecolor{shadowcolor}{rgb}{0.0000, 0.0000, 0.6179}  
\definecolor{bulletcolor}{rgb}{ 0.8441, 0.1582, 0.0000} 
\definecolor{deepskyblue}{rgb}{0, 0.75, 1.0}
\definecolor{royalblue}{rgb}{0.254901960784314,   0.411764705882353,   0.882352941176471}
\definecolor{dodgerblue}{rgb}{0.11765, 0.56471, 1.0}
\definecolor{bordercolor}{rgb}{0,0,.2380}               

\definecolor{lightskyblue}{rgb}{0.53, 0.81, 0.98}
\definecolor{magenta}{rgb}{1.0, 0.0, 1.0}
\definecolor{lavendermagenta}{rgb}{0.93, 0.51, 0.93}
\definecolor{internationalorange}{rgb}{1.0, 0.31, 0.0}

\newcommand{\ansA}[1]{{\color{black}#1}}
\newcommand{\ansB}[1]{{\color{black}#1}}
\newcommand{\ansBB}[1]{{\color{black}#1}}

\title[Modelling uncertainty in a 2-layer QG model] 
      {Modelling uncertainty using stochastic transport noise in a 2-layer quasi-geostrophic model}

\author[Colin Cotter, Dan Crisan, Darryl Holm, Wei Pan, Igor Shevchenko]{}


 \email{colin.cotter@imperial.ac.uk}
 \email{d.crisan@imperial.ac.uk}
 \email{d.holm@imperial.ac.uk}
 \email{wei.pan@imperial.ac.uk}
 \email{i.shevchenko@imperial.ac.uk}

\thanks{$^*$ Corresponding author: Igor Shevchenko}

\begin{document}
\maketitle

%

\centerline{\scshape Colin Cotter, Dan Crisan, Darryl Holm, Wei Pan, Igor Shevchenko$^*$}
\medskip
{\footnotesize
 \centerline{Department of Mathematics, Imperial College London}
   \centerline{Huxley Building, 180 Queen's Gate}
   \centerline{London, SW7 2AZ, UK}
}

\bigskip

 \centerline{(Communicated by the associate editor name)}

\begin{abstract}
The stochastic variational approach for geophysical fluid dynamics was
introduced by Holm (Proc Roy Soc A, 2015) as a framework for deriving
stochastic parameterisations for unresolved scales. 
This paper applies the variational stochastic parameterisation in a two-layer
quasi-geostrophic model for a $\beta$-plane channel flow configuration. 
We present a new method for
estimating the stochastic forcing (used in the parameterisation) to approximate unresolved components using
data from the high resolution deterministic simulation, and 
describe a procedure for computing physically-consistent initial conditions for the stochastic model.
We also quantify uncertainty of coarse grid simulations relative to the fine grid ones
in homogeneous (teamed with small-scale vortices) and 
heterogeneous (featuring horizontally elongated large-scale jets) flows, and analyse how the spread of stochastic solutions 
depends on different parameters of the model. 
The parameterisation is tested by comparing it with the true eddy-resolving
solution that has reached some statistical equilibrium and the deterministic solution
modelled on a low-resolution grid.
The results show that the proposed parameterisation
significantly depends on the resolution of the stochastic model and 
gives good ensemble performance for both 
homogeneous and heterogeneous flows, and \ansA{the parameterisation lays} solid foundations for data assimilation.
\end{abstract}

\section{Introduction}

\subsection{Motivation}


\ansA{
The process of ``upscaling", or ``coarse graining'' of the fine-scale data is in common usage in computational simulations at coarser scales. 
The goal of the present paper is to quantify the uncertainty in the process of upscaling, or coarse graining of fine-scale computationally simulated data for use in computational simulations at coarser scales, in the example of a two-level quasigeostrophic channel flow.} 

\ansA{The question for coarse graining that we address in this paper is the following:
\emph{How can we use computationally simulated surrogate data at well resolved scales, in combination with the mathematics of stochastic processes in nonlinear dynamical systems, to estimate and model the effects on the simulated variability at much coarser scales of the computationally unresolvable, small, rapid, scales of motion at the finer scales?} We will address this question in the context of two-level quasigeostrophic channel flow.}

\ansBB{
Stochastic parameterisation has been a very active research topic in the last 20 years and 
a number of different approaches have been proposed.
\cite{Majda_et_al2001,franzke2005low} developed a general approach based on applying singular perturbation
theory to the Fokker-Planck equation for the dynamical system. 
\cite{wouters2012disentangling} proposed a rather general approach
based on linear response theory that the system being modelled is
formed from weak coupling between two self-contained dynamical
systems, and leads to the introduction of a stochastic term plus a
memory term. A parameterisation was proposed in~\cite{Abramov2016}
based on averaging a system of slow and fast variables over the
invariant measure of the slow variables given fixed values of the fast
ones; this idea was investigated within a low-order coupled
ocean-atmosphere model in~\cite{vannitsem2014stochastic}.

In ocean modelling there also exists a tradition of ``stochastic backscatter’’ which goes back to~\cite{Leith1990}, who first developed the 
stochastic backscatter prescription is as a supplement to the Smagorinsky viscosity and applied it in simulations of a planar turbulent shear mixing layer. 
Since then, the stochastic backscatter prescription has become a standard feature for modelling eddy effects in ocean modelling. See, e.g.,~\cite{PortaMana_Zanna2014} 
for a good treatment of the history of this development. The stochastic backscatter prescription differs from the current approach in 
that it represents stochastic forcing in the PV equation, rather than stochastic transport. 

Our approach, known as Stochastic Advection by Lie Transport (SALT), differs from these methods in that its
slow/fast decomposition is derived at the level of the Lagrangian
flow map for the fluid, with a stochastic Ansatz being derived for the
flow map in the diffusive limit~\cite{CoGoHo2017}. This Anzatz simply states that fluid particles move with the
mean flow plus a stochastic perturbation to the velocity that comes
from averaging the small-scale dynamics. Then, the Ansatz can be
passed to a variational principle for geophysical fluid dynamics
equations, leading to Eulerian SPDEs where the stochastic terms appear
as modifications to the transport terms.

The Ansatz for a stochastic Lagrangian flow map goes back to~\cite{Kraichnan1968}. It was also the basis for the approach of 
Location Uncertainty (LU) advocated in~\cite{Memin2014}. However, the application of Lagrangian transport noise in the LU model differs 
from that in the SALT model, and both the LU model and the SALT model differ from Kraichnan’s advection model. 
Now, Kraichnan’s model applies transport noise to advection of a scalar fluid quantity. The LU model applies stochastic scalar 
transport to the momentum in the fluid motion equation. The SALT model applies stochastic scalar advection as a constraint on 
the back-to-labels map in Hamilton’s variational principle for the fluid motion equation in~\cite{holm2015variational}. 
As result of the different interpretations of Kraichnan’s stochastic scalar model, the stochastic Euler fluid equations in the LU model 
and in the SALT have mutually exclusive conservation laws.  Namely, the LU fluid motion equation preserves kinetic energy and not Kelvin 
circulation, while the SALT fluid motion equation preserves Kelvin circulation and not kinetic energy.  
}

Our approach is guided by recent results in  \cite{CoGoHo2017} which showed that a multi-scale decomposition of the deterministic Lagrange-to-Euler fluid flow map $g_t$ into a slow large-scale mean and a rapidly fluctuating small-scale map leads to Lagrangian fluid paths $x_t=g_tX$ with $g_0=Id$ on a manifold $ \mathcal{D} $ governed by the stochastic process $g_t\in {\rm Diff}(\mathcal{D})$ on the Lie group of diffeomorphic flows, which appears in the same form as had been proposed and studied for fluids in \cite{holm2015variational}; namely,
\begin{equation}
{\dd}x_t  = {\dd}g_t \,X
= u_t(x)dt + {\color{red}\sum\limits^K_{k=1} \xi^k(x)\circ dW^k_t}
=   u_t(g_tX)dt + {\color{red}\sum\limits^K_{k=1} \xi^k(g_tX)\circ dW^k_t} 
\,,\label{Lag-stoch-process}
\end{equation}
where $x=g_tX$, ${\dd}$ represents stochastic differentiation, the divergence-free vector fields 
$\xi^{k}(x)=\hat{z}\times\nabla\zeta^{k}(x)$ for $k=1,2,\dots,K,$ \ansA{where the $\zeta^k(x)$ are the streamfunctions for the prescribed vector functions $\xi^k(x)$
(to be described below)}, 
$\hat{z}$ denotes the z-axis,
$x\in \mathcal{D}$ on the domain of flow $\mathcal{D}$, 
and $\circ\, dW^{k}(t)$ denotes the Stratonovich differential with independent Brownian motions 
$dW^{k}(t)$. The stochastic process for the evolution of the Lagrangian process $g_t$ 
in equation \eqref{Lag-stoch-process} involves the pullback of the Eulerian total velocity vector field, 
which comprises the sum of a drift displacement vector field $u_t(x)dt$ plus a sum over terms in 
$\xi^{k}(x)$ representing the (assumed stationary) spatial correlations of the temporal noise in the Stratonovich representation, each with its own independent Brownian motion in time.
\ansA{The $\xi$'s represent spatial correlations of the rapid fluctuations which when replaced by noise in 
the homogenisation limit provide a vector of stochastic transport at the spatial coarse scale of the slow variables. 
This extra transport introduces uncertainty and widens the ensemble spread of the flow rather than 
simply improving the mean.}

The choice of the Stratonovich form (as opposed to the It\^{o} form) in the differential equation ensures 
that the variational derivation of SPDE makes sense. 
This is because the Stratonovich form satisfies the standard chain rule making the formal computation straightforward. 
Whilst the Stratonovich form is preferable from a physical/modelling perspective, 
the rigorous analysis of the solution of the SPDE requires the transformation from the Stratonovich 
form to the It\^{o} form through the addition of a covariation between the integrand and the driving Brownian motion. 
Further details of the Stratonovich/It\^{o} integral can be found, for example, in~\cite{KaratzasShreve1991}.

In \cite{holm2015variational} the velocity decomposition formula \eqref{Lag-stoch-process} was applied in the Hamilton-Clebsch variational principle to derive coadjoint motion equations as stochastic partial differential equations (SPDEs) whose ensemble of realisations  can be used to quantify the uncertainty in the slow dynamics of the resolved mean velocity $u_t(x)$.  Under the conditions imposed 
in the derivation of formula \eqref{Lag-stoch-process} in \cite{CoGoHo2017} using homogenization theory, the sum of vector fields in \eqref{Lag-stoch-process} that had been treated in \cite{holm2015variational} from the viewpoint of stochastic coadjoint motion was found to represent a bona fide decomposition of the fluid transport velocity into a mean plus fluctuating flow.

\subsection{Data-driven modelling of uncertainty} 
The most familiar example of data-driven modelling occurs in numerical weather prediction (NWP). In NWP, various numerically unresolvable, but observable, local subgrid-scale processes, such as formation of fronts and generation of tropical cyclones, are expected to have profound effects on the variability of the weather. These subgrid-scale processes must be parameterized at the resolved scales of the numerical simulations. Of course, the accuracy of a given parameterization model often remains uncertain. In fact, even the possibility of modelling subgrid-scale properties in terms of resolved-scale quantities available to simulations may sometimes be questionable. However, if some information about the \textit{statistics} of the small-scale excitations is known, such as the spatial correlations of its observed transport properties at the resolved scales, 
one may arguably consider modelling the effects of the small scale dynamics on the resolved scales by a stochastic transport process whose spatial correlations match the observations, at the computationally unresolvable scales. 
The $\xi$ functions are the leading EOFs of the differences between the fine-grained and coarse-grained Lagrangian trajectories, evaluated on the fixed coarse grid.

\subsection{The main content of the paper}
The rest of the paper is structured as follows. 
Section \ref{sec:2d_qg} presents the deterministic and stochastic QG equations.
Section \ref{sec:2d_qg_num} is divided into two parts. In the first part we present deterministic
simulations of heterogeneous and homogeneous flows at different resolutions.
In the second part, we develop a method for calibrating the $\xi$'s by using the Lagrangian description of the flow. 
We also present a procedure for computing physically-consistent initial conditions for the stochastic model.
Then, we quantify uncertainty in the homogeneous and heterogeneous solutions, and analyse how the spread of stochastic solutions 
depends on different parameters of the model. In Section~\ref{sec:deter_vs_stoch} we compare the true solution, 
the deterministic solution modelled on the low-resolution grid, and the parameterised solution.
In Section~\ref{sec:concl} we provide conclusions and outlook for future research.

\section{The two-dimensional multilayer quasi-geostrophic model\label{sec:2d_qg}}
\subsection{Deterministic case}
\begin{figure}[htp]
\begin{center}
\includegraphics[scale=0.15]{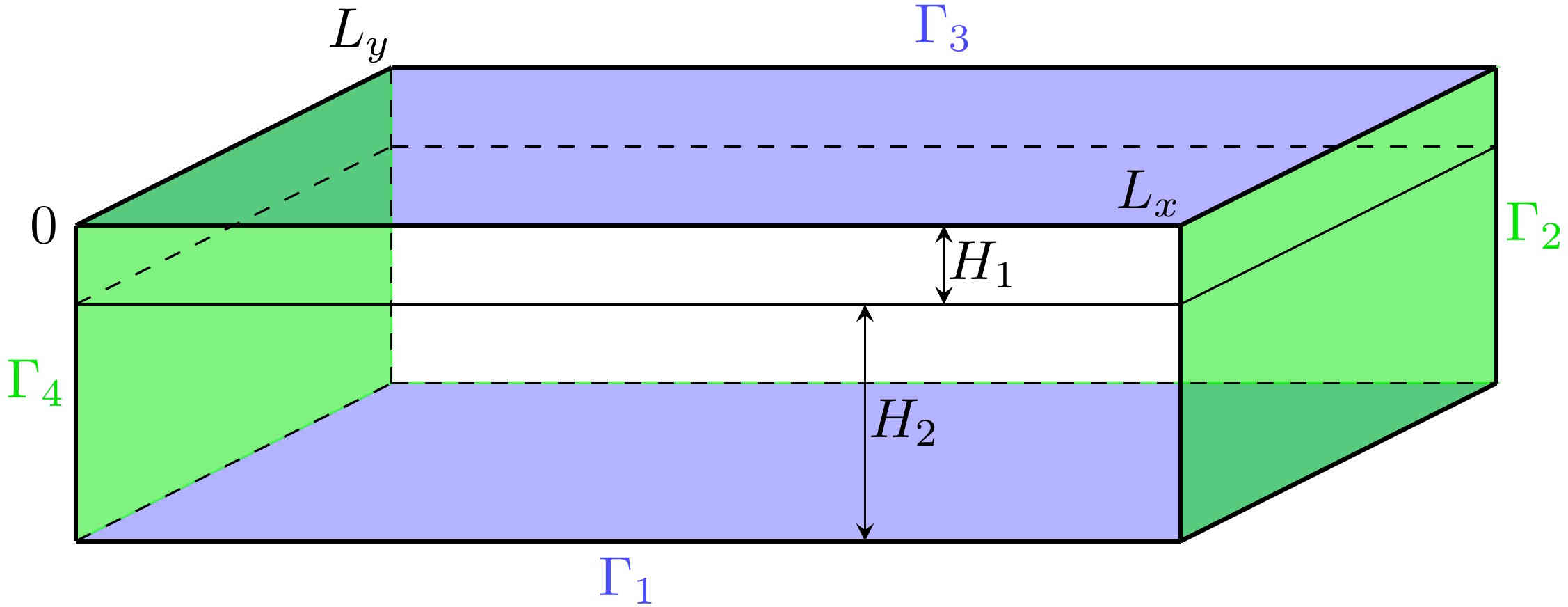}
\caption{Shown is a schematic of the computational domain $\Omega$.}
\label{fig:gem_setup}
\end{center}
\end{figure}

The two-layer deterministic QG equations for the potential vorticity (PV) anomaly $q$ in a 
domain $\Omega$ are given by the PV material conservation law augmented with forcing and dissipation~\cite{Pedlosky1987,Vallis2006}:
\begin{subequations}
\begin{align}
\partial_t q_1+J(\psi_1,q_1+\beta y)&=\nu\Delta^2\psi_1,\\
\partial_t q_2+J(\psi_2,q_2+\beta y)&=\nu\Delta^2\psi_2-\mu\Delta\psi_2,
\end{align}
\label{eq:pv}
\end{subequations}
where $\psi$ is the stream function, $J(f,g)=f_xg_y-f_yg_x$ is the Jacobian, the planetary vorticity gradient is given 
by parameter $\beta$, $\mu$ is the bottom friction parameter, and $\nu$ is the eddy viscosity.
The computational domain $\Omega=[0,L_x]\times[0,L_y]\times[0,H]$ is a horizontally periodic flat-bottom channel of depth
$H=H_1+H_2$ given by two stacked isopycnal fluid layers of depth
$H_1$ and $H_2$, respectively (Figure~\ref{fig:gem_setup}). 

Forcing in system~\eqref{eq:pv} is introduced through a vertically sheared, baroclinically unstable background flow (e.g.,~\cite{BerloffKamenkovich2013})
\begin{equation}
\psi_i\rightarrow-U_i\,y+\psi_i,\quad i=1,2, 
\label{eq:forcing}
\end{equation}
where the parameters $U_i$ are background-flow zonal velocities.

The PV anomaly and stream function are related through two elliptic equations:
\begin{subequations}
\begin{align}
q_1=\Delta\psi_1+s_1\psi_{[21]},\\
q_2=\Delta\psi_2+s_2\psi_{[12]},
\end{align}
\label{eq:q_psi}
\end{subequations}
with stratification parameters $s_1$, $s_2$; $\psi_{[ij]}:=\psi_i-\psi_j$.

System~(\ref{eq:pv})-(\ref{eq:q_psi}) is augmented by the integral mass conservation constraint~\cite{McWilliams1977}
\begin{equation}
\partial_t\iint\limits_{\Omega}\psi_{[12]}\ dydx=0,
\label{eq:masscon}
\end{equation}
as well as by the periodic horizontal boundary conditions,
\begin{equation}
\boldsymbol{\psi}\Big|_{\Gamma_2}=\boldsymbol{\psi}\Big|_{\Gamma_4}\,,\quad \boldsymbol{\psi}=(\psi_1,\psi_2)\,,
\label{eq:bc24}
\end{equation}
and no-slip boundary conditions at the lateral (northern and southern) boundaries of the channel,
\begin{equation}
\boldsymbol{u}\Big|_{\Gamma_1}=\boldsymbol{u}\Big|_{\Gamma_3}=0\,,
\label{eq:bc13}
\end{equation}
where $\boldsymbol{u}=(u,v)$ is the velocity vector, and $\Gamma_1$, $\Gamma_2$, $\Gamma_3$, $\Gamma_4$ are the boundaries
of the computational domain (Figure~\ref{fig:gem_setup}).

The QG model~\eqref{eq:pv}-\eqref{eq:bc13} is solved using the high-resolution CABARET method, which is  based
on a second-order, non-dissipative and low-dispersive, conservative advection scheme~\cite{Karabasov_et_al2009}.
The distinctive feature of this scheme is its ability to simulate large-Reynolds-number flow regimes at much lower
computational costs compared to conventional methods (see, e.g.,~\cite{Arakawa1966,WoodwardColella1984,ShuOsher1988,Hundsdorfer_et_al1995}),
since the scheme is low dispersive and non-oscillatory, 
it has a compact computational stencil in both space and time, and no dissipation error.
We refer the reader to the study~\cite{Karabasov_et_al2009} for more detailed discussion of the CABARET method.
\ansA{The main salient feature here is that the CABARET method runs stably without diffusion, since the flux limiting scheme is sufficient to absorb enstrophy at small scales. We include diffusion in the model here purely to maintain a statistical steady state as required for our 
stochastic parameterisation.}

\subsection{Stochastic case\label{sec:SQG}}
The stochastic version of the two-layer QG equations is given by the following system of equations~\cite{holm2015variational}:
\begin{align}
\label{eq:SLTpv}
\begin{split}
\dd q_1+J(\psi_1\,dt +{\color{red} \sum\limits^K_{k=1} \zeta^k_1 \circ dW^k_t},\, q_1+\beta y)
&=(\nu\Delta^2\psi_1)\,dt
,\\
\dd q_2+J(\psi_2\,dt +{\color{red} \sum\limits^K_{k=1} \zeta^k_2 \circ dW^k_t}, \, q_2+\beta y)
&=(\nu\Delta^2\psi_2-\mu\Delta\psi_2)\,dt.
\end{split}
\end{align}

The stochastic terms marked in red color are the only
difference from the deterministic QG model~\eqref{eq:pv}. All other equations remain the same as in the deterministic case.
As explained in the Introduction, the noise is Gaussian, white in time, with zero mean and its velocity-velocity correlation function may be written in terms of its eigenvectors $\xi^k=\hat{z}\times\nabla\zeta^k$. This provides an
interpretation of the $\xi^k$.

\section{Numerical results\label{sec:2d_qg_num}}

\subsection{Deterministic case}
We define the computational domain $\Omega=[0,L_x]\times[0,L_y]\times[0,H]$ as a horizontally periodic flat-bottom channel with $L_x=3840\, \rm km$, $L_y=L_x/2\, \rm km$, and total depth
$H=H_1+H_2$, with $H_1=1.0\, \rm km$, $H_2=3.0\, \rm km$ (Figure~\ref{fig:gem_setup}). 
We choose governing parameters of the QG model that are typical to a mid-latitude setting.
These comprise the planetary vorticity gradient $\beta=2\times10^{-11}\, {\rm m^{-1}\, s^{-1}}$, eddy viscosity $\nu=3.125\,\rm m^2 s^{-1}$, and
the bottom friction parameters $\mu=\{4\times10^{-8},4\times10^{-7}\}\, {\rm s^{-1}}$. We will explain this choice below, as well as the reason for studying two different flow regimes.
The background-flow zonal velocities in~\eqref{eq:forcing} are given by $U=[6.0,0.0]\,\rm cm\, s^{-1}$, while
the stratification parameters in system~\eqref{eq:q_psi} are $s_1=4.22\cdot10^{-3}\,\rm km^{-2}$, $s_2=1.41\cdot10^{-3}\,\rm km^{-2}$; chosen so that the first Rossby deformation radius is $Rd_1=25\, {\rm km}$.
In order to ensure that the numerical solutions are statistically equilibrated, the model is initially spun up from the state of rest to $t=0$ over the time
interval $T_{spin}=[-100,0]\, {\rm years}$. 

\begin{figure}[htp]
\begin{center}
\includegraphics[scale=0.175]{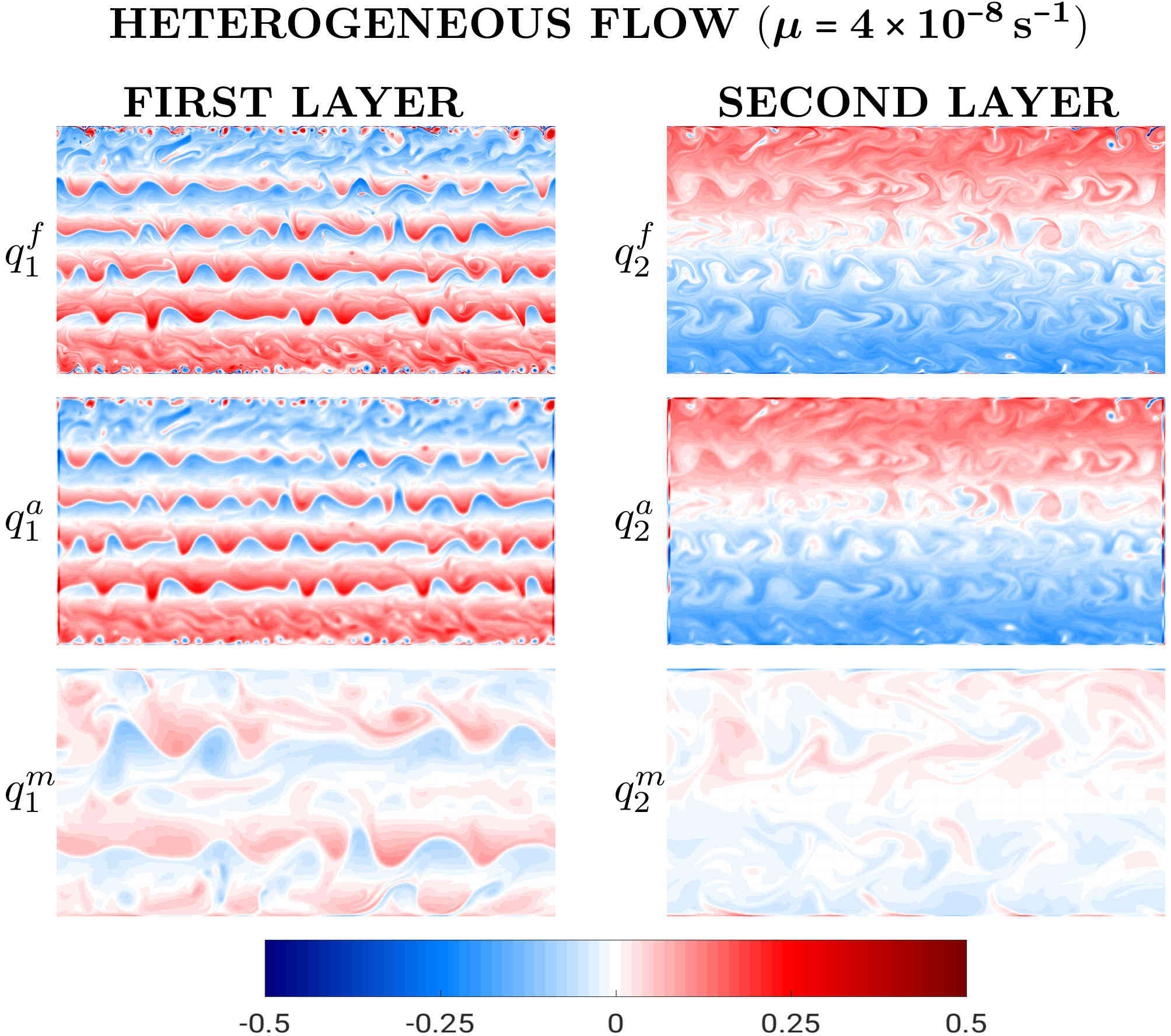}

\caption{The series of snapshots shows the high-resolution solution $q^f$ computed on the fine grid $G^f=2049\times1025$ ($dx\approx dy\approx 1.9\, {\rm km}$),
the true solution $q^a$ computed on the coarse grid $G^c=257\times129$ ($dx\approx dy\approx 15\, {\rm km}$), and 
the low-resolution solution $q^m$ also computed on $G^c$ by simulating the QG model 
for the \textit{\textbf{low drag}} $\boldsymbol{\mu=4\times10^{-8}\, {\rm s^{-1}}}$ (\textbf{\textit{heterogeneous flow}}).
All the solutions are given in units of $[s^{-1}f^{-1}_0]$,  where $f_0=0.83\times10^{-4}\, {\rm s^{-1}}$ is the Coriolis parameter.
In order to visualize the solutions on the same color scale we have multiplied the ones in the second layer by a factor of 5.}
\label{fig:qf_qa_qc_qam_qcm_mu1D-8_257x129}
\end{center}
\end{figure}

\begin{figure}[htp]
\begin{center}
\includegraphics[scale=0.175]{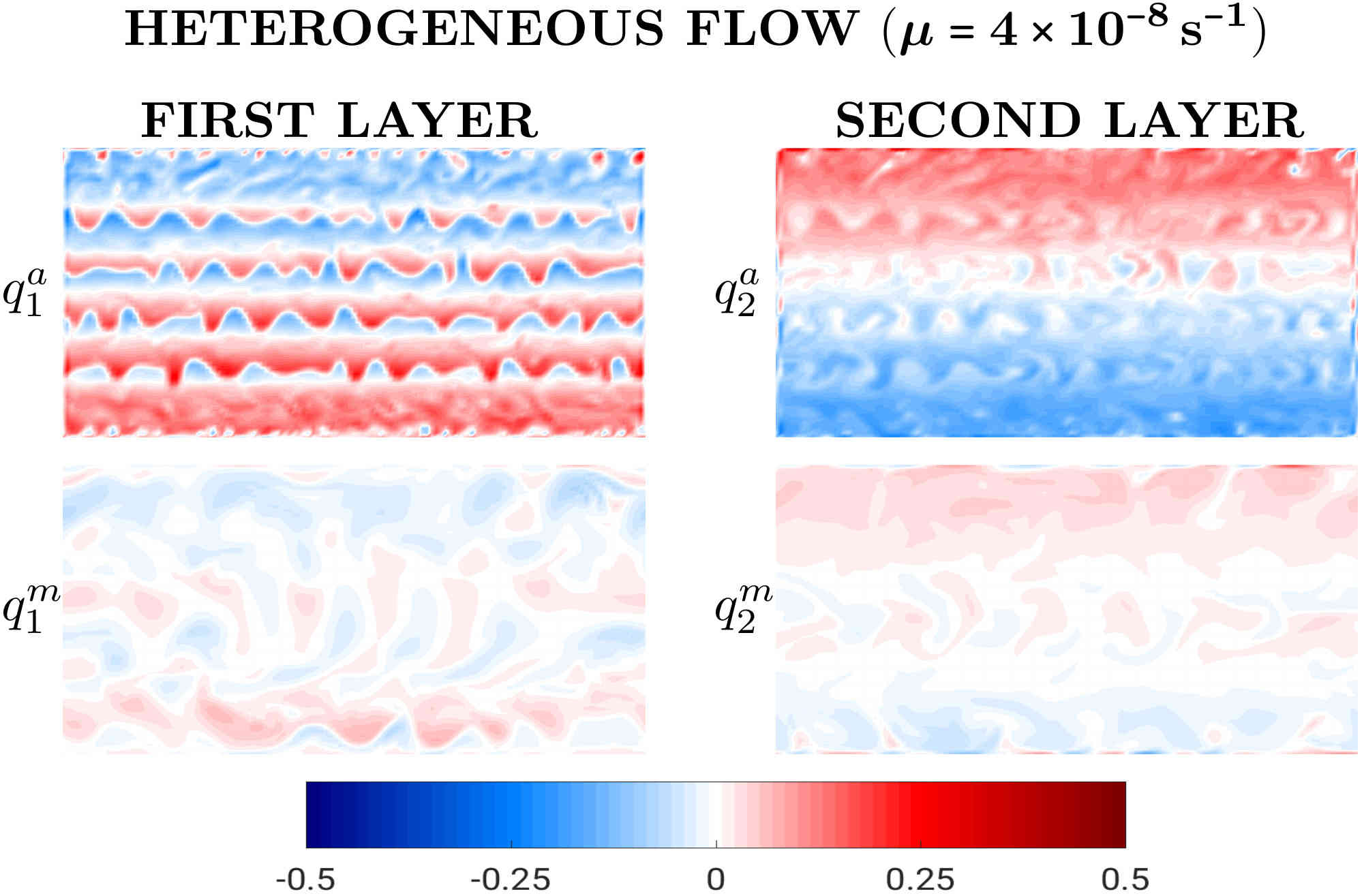}


\caption{The same as in Figure~\ref{fig:qf_qa_qc_qam_qcm_mu1D-8_257x129}, but for $G^c=129\times65$ ($dx\approx dy\approx 299\, {\rm km}$).
}
\label{fig:qf_qa_qc_qam_qcm_mu1D-8_129x65}
\end{center}
\end{figure}

\begin{figure}[htp]
\begin{center}
\includegraphics[scale=0.175]{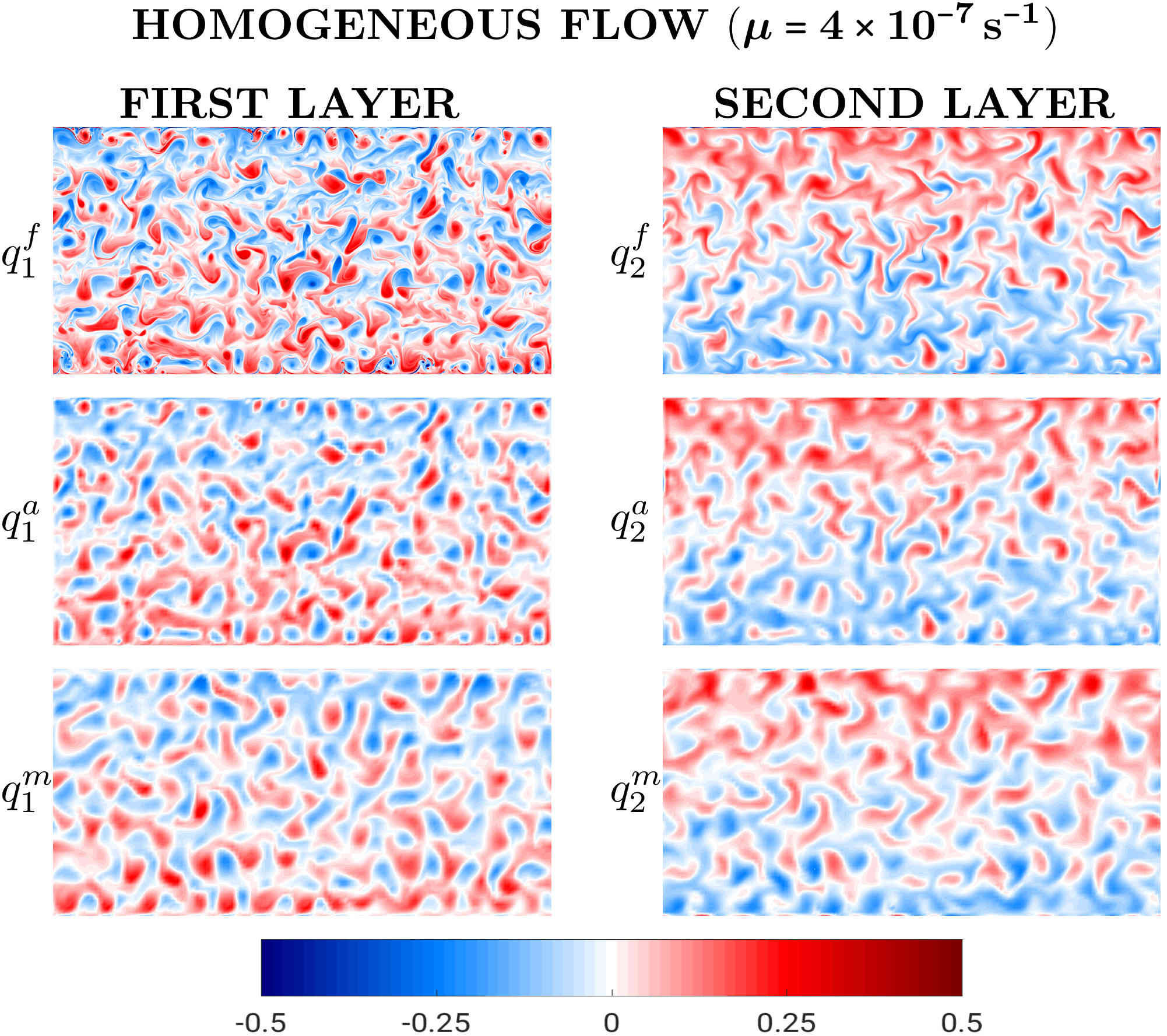}

\caption{The series of snapshots in the figure shows the high-resolution solution $q^f$ computed on the fine grid $G^f=2049\times1025$ ($dx\approx dy\approx 1.9\, {\rm km}$),
the true solution $q^a$ computed on the coarse grid $G^c=129\times65$ ($dx\approx dy\approx 299\, {\rm km}$), and 
the low-resolution solution $q^m$ computed on the coarse grid $G^c$ by simulating the QG model for the \textit{\textbf{high drag}}
$\boldsymbol{\mu=4\times10^{-7}\, {\rm s^{-1}}}$ (\textbf{\textit{homogeneous flow}}).
All the solutions are given in units of $[s^{-1}f^{-1}_0]$,  where $f_0=0.83\times10^{-4}\, {\rm s^{-1}}$ is the Coriolis parameter.
In order to visualize the solutions on the same color scale we have multiplied the ones in the second layer by a factor of 5.}
\label{fig:qf_qa_qc_qam_qcm_mu1D-7}
\end{center}
\end{figure}

For the purpose of this paper we consider a heterogeneous (Figures~\ref{fig:qf_qa_qc_qam_qcm_mu1D-8_257x129} and~\ref{fig:qf_qa_qc_qam_qcm_mu1D-8_129x65}) 
and homogeneous (Figure~\ref{fig:qf_qa_qc_qam_qcm_mu1D-7}) flow regime 
(which correspond to flows with low ($\mu=4\times10^{-8}\,\rm{s^{-1}}$) and high ($\mu=4\times10^{-7}\,\rm{s^{-1}}$) drag, 
respectively) at different resolutions, and study how the parameterisation performs in each case.
In brief,  for the smaller bottom friction, we find that jet-like structures (also referred to as striations) emerge in high-resolution
simulations, resulting from interplay of forcing, damping, and baroclinic instability. 
For larger bottom friction, the flow pattern is essentially homogeneous and no coherent jet-like structures are seen in this case.

Before presenting the results, it is helpful to explain the following four
solutions used throughout the paper. In particular, 
\begin{itemize}
 \setlength\itemsep{0.0cm}
 \item High-resolution deterministic solution $q^f$ computed on the fine grid 
       $G^f=\{N_x\times N_y\}$, $N_x=2049$, $N_y=1025$  ($dx\approx dy\approx 1.9\, {\rm km}$).
       This solution is only used to compute the true solution, $q^a$, described below;
 \item Low-resolution deterministic solution $q^a$ computed on the coarse grids 
       $G^c=257\times129, dx\approx dy\approx 15\, {\rm km}$ (Figure~\ref{fig:qf_qa_qc_qam_qcm_mu1D-8_257x129}, middle row) and 
       $G^c=129\times65, dx\approx dy\approx 299\, {\rm km}$ (Figure~\ref{fig:qf_qa_qc_qam_qcm_mu1D-8_129x65}, upper row) 
       as the solution of the elliptic equation~\eqref{eq:q_psi} with the stream function $\psi^a$, where $\psi^a$ is computed
       by spatially averaging the high-resolution stream function $\psi^f$ over the coarse grid cell $G^c$.
       We refer to $q^a$ as \textit{the truth} or \textit{the true solution}, and use it for comparison with the parameterised solution.
       Moreover, solving the QG equations on the low resolution grid (say, $G^c=257\times129$) gives the solution 
       (Figure~\ref{fig:qf_qa_qc_qam_qcm_mu1D-8_257x129}, bottom row)
       which has nothing in common with the truth (Figure~\ref{fig:qf_qa_qc_qam_qcm_mu1D-8_257x129}, middle row), 
       not to mention the high-resolution solution (Figure~\ref{fig:qf_qa_qc_qam_qcm_mu1D-8_257x129}, top row). The same is true
       for the coarser grid $G^c=129\times65$ (compare the upper and second row in Figure~\ref{fig:qf_qa_qc_qam_qcm_mu1D-8_129x65}).
       Note that we do not introduce extra notations to differentiate between $q^a$ computed on different grids, since 
       we will discuss each case separately;
 \item Low-resolution modelled solution $q^m$ (also referred to as the coarse-grid modelled solution) computed on the coarse grids 
       $G^c=\{257\times129, 129\times65\}$ by simulating the QG model. This solution is used for parameterisation. 
 \item Low-resolution parameterised solution $q^p$ (also referred to as the stochastic solution) computed on the coarse grids 
       $G^c=\{257\times129, 129\times65\}$ by simulating the stochastic QG model.
\end{itemize}

\begin{figure}[htp]
\begin{center}
\includegraphics[scale=0.175]{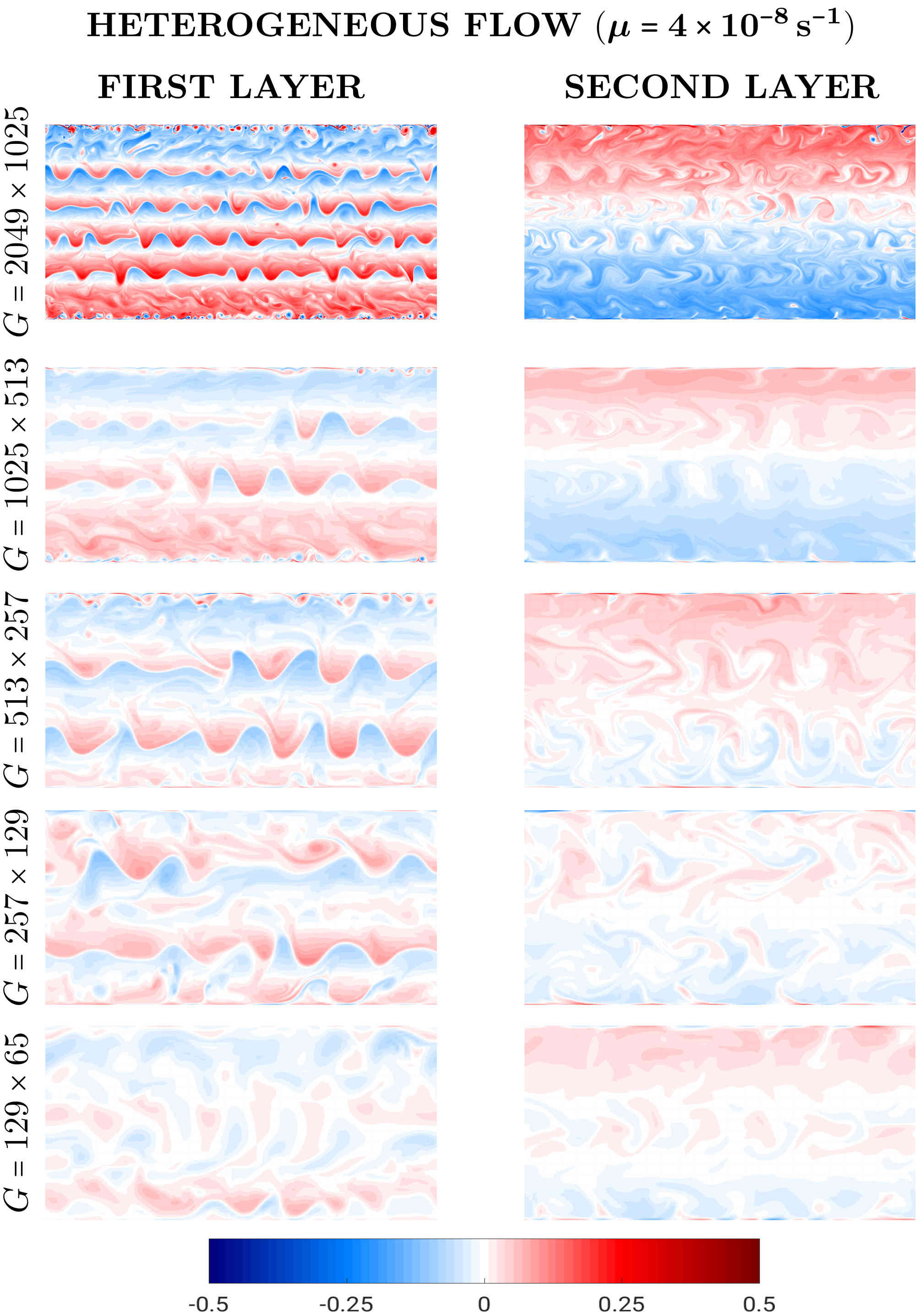}

\caption{The series of snapshots shows the dependence of the solution on the resolution
for the \textit{\textbf{low drag}} $\boldsymbol{\mu=4\times10^{-8}\, {\rm s^{-1}}}$ (\textbf{\textit{heterogeneous flow}}).
All the solutions are given in units of $[s^{-1}f^{-1}_0]$,  where $f_0=0.83\times10^{-4}\, {\rm s^{-1}}$ is the Coriolis parameter.
In order to visualize the solutions on the same color scale we have multiplied the ones in the second layer by a factor of 5.}
\label{fig:qf_all_grids}
\end{center}
\end{figure}

For the low drag, which corresponds to the bottom friction coefficient $\mu=4\times10^{-8}$, flow dynamics is highly-heterogeneous,
and small-scale features are prevalent in the first layer 
(Figures~\ref{fig:qf_qa_qc_qam_qcm_mu1D-8_257x129} and~\ref{fig:qf_qa_qc_qam_qcm_mu1D-8_129x65}).
The high-resolution flow $q^f$ (computed on the fine grid $G^f=2049\times1025$ with the resolution $dx\approx dy\approx 1.9\, {\rm km}$) in the first layer consists of two flow regions: 
the fast flow within the jets and the slow flow between the jets. 
The flow dynamics in the first layer teems with small-scale eddies which, in turn, 
maintain the striated structure of the flow (see, e.g.~\cite{Kamenkovich_et_al2009}).
The flow dynamics in the second layer is much less energetic than that of the first one, and exhibits
neither small-scale features nor jets. 
The true solution $q^a$, computed on the coarse grids
$G^c=\{257\times129, 129\times65\}$, captures the striated flow structure, small-scales features
as well as flow energetics of the high-resolution solution $q^f$ in the first and second layer.
However, the low-resolution modelled solution $q^m$ (the solution which has to be parameterised and then used in uncertainty
quantification tests presented in Section~\ref{sec:xi_SQG}) computed on the coarse grid $G^c$ by simulating the QG model
cannot capture the correct structure (the number of jets and their positions) of the true flow dynamics. 
For example, $q^m_1$ on the grid $257\times129$ has only two jets 
(Figure~\ref{fig:qf_qa_qc_qam_qcm_mu1D-8_257x129}, bottom row), while the true solution 
$q^a_1$ (Figure~\ref{fig:qf_qa_qc_qam_qcm_mu1D-8_257x129}, middle row) has four jets; we only count jets within the domain and 
exclude boundary layers from the consideration.
The situation on the coarse grid $129\times65$ is even worse: the low-resolution solution $q^m_1$ 
(Figure~\ref{fig:qf_qa_qc_qam_qcm_mu1D-8_129x65}, bottom row) has only one very latent jet if at all.
Thus, the coarse-grid QG equations fail to model the proper jet-like structure of the flow. Apparently, 
the coarse resolution suppresses the small-scale eddies, which are thought to be  one of the mechanisms 
responsible for maintaining the jets (see, e.g.~\cite{Kamenkovich_et_al2009}). 
In fact, we have found that the solution bifurcates between 
the grid $2049\times1025$ (four jets~\footnote{\ansA{We have checked that higher resolutions (up to $4097\times2049$) do not lead to stronger striations and more jets. Using even higher resolutions is beyond our hardware capabilities.}}) and $1025\times513$ (only two jets), see Figure~\ref{fig:qf_all_grids}. 
In other words, there is no continuous degradation of coherent structures as the resolution decreases like, for example,
in the double-gyre problem (e.g.~\cite{SB2015}).
Instead, one observes a fast switch from a solution with four jets to a solution 
with two jets, and eventually to a solution with only one weak, blurred  jet on the grid $129\times65$ (Figure~\ref{fig:qf_all_grids}). 
This bifurcation of the solution makes it very difficult to parameterise, 
since the low-resolution solution has a completely different structure compared to the true solution.
%
%
%
%

Going a bit ahead, we would like to note that our analysis of the time averaged total energy, $\bar{E}$ of the solutions computed on different grids
$(\bar{E}^{2049\times1025}_{\rm heter}=26$, $\bar{E}^{1025\times513}_{\rm heter}=29$, $\bar{E}^{513\times257}_{\rm heter}=28$,
$\bar{E}^{257\times129}_{\rm heter}=26$, $\bar{E}^{129\times65}_{\rm heter}=3)$
shows a sudden one order of magnitude drop between the grids $\bar{E}^{257\times129}_{\rm heter}$ and $\bar{E}^{129\times65}_{\rm heter}$;
here, the energy is non-dimensional.
This observation will be important later on for explaining why the parameterisation for the coarse grid $129\times65$
works worse than for the finer grid $257\times129$.

Getting back to the analysis of QG solutions,
we have found that for the high drag, with bottom friction coefficient $\mu=4\times10^{-7}$, the 
flow dynamics becomes more homogeneous (Figure~\ref{fig:qf_qa_qc_qam_qcm_mu1D-7}).
As in the heterogeneous case, the high-resolution flow $q^f$ is 
more energetic in the first layer than in the second, although 
this difference in the structure of the flow is less pronounced than that in the heterogeneous case. 

Comparing the high-resolution solution $q^f$ with its coarse-grid analogue $q^a$ we conclude that the latter
captures the small-scales features as well as the energetics of $q^f$ ($\bar{E}^f_{\rm homog}=26$) 
in both layers; the time-averaged total energy of the space averaged solution is $\bar{E}^a_{\rm homog}=21$.
Unlike the heterogeneous flow, the coarse-grid modelled solution $q^m$ (the solution we parameterise and use in uncertainty
quantification tests given in Section~\ref{sec:xi_SQG}) is also homogeneous in structure and adequately restores the energetics of 
the true solution $q^a$; the time-averaged total energy of the modelled solution is $\bar{E}^m_{\rm homog}=17$.
In other words, the coarse-grid QG model properly represents the large-scale flow dynamics for flows with
higher drag. In the case of high-drag flows,
the zonally uniform eigenmodes responsible for maintaining the jet-like structure of the flow  
become more damped thus making jets much more latent compared with low-drag flows (see, e.g.~\cite{Berloff_et_al2011}). 
The results presented in this section show that more
interesting and energetic flow dynamics is confined in the first layer. 
Therefore, from now on we will focus our attention on the first layer unless stated otherwise.

\subsection{Stochastic case\label{sec:2d_qg_num_stochastic}}

\subsubsection{Calibration of eigenvectors~\label{sec:cal}}

  We approach the choice of $\xi_i$'s by starting from the
  Ansatz that models the effect of small scales on Lagrangian particles by noise, and which underpins
  the derivation of the stochastic QG model.
  We
present a new methodology for modelling the difference between passive Lagrangian particles advected 
by the high-resolution deterministic velocity field $\bf u$ computed on the fine grid $G^f=2049\times1025$ and its coarsened counterpart 
$\bar{\bf u}$ computed on the coarse grid $G^c=129\times65$ by differentiating the coarse-grained stream function $\psi^a$. 
The stream function is
computed by spatially averaging the high-resolution stream function $\psi^f$ over the coarse grid cell $G^c$.
Based on this difference, we compute Empirical Orthogonal Functions (EOFs)~\cite{Preisendorfer1988,HaJoSt2007}, which are then 
used as the $\xi$'s in the parameterised QG model.
We also perform uncertainty quantification tests for the stochastic differential equation for Lagrangian particles~\eqref{eq:barx} and the stochastic QG model~\eqref{eq:SLTpv},
and study how the number of EOFs and size of the ensemble of stochastic solutions (referred to as ensemble members) affect the width of the stochastic spread.
Note that we present the results only for the heterogeneous flow on the grid $G^c=129\times65$, 
since the goal of this section is to demonstrate how the methodology works.

\subsubsection{Measuring the Lagrangian evolution}
In the stochastic GFD framework, stochastic PDEs are derived from the
starting assumption that (coarse-grained) fluid particles satisfy the
equation
\begin{equation}
  \label{eq:barx}
\diff \bar{\bf x}(a,t) = \bar{\bf u}(\bar{\bf x}(a,t),t)\diff t + {\color{red}\sum\limits^K_{k=1} \xi^k(\bar{\bf x}(a,t)) \circ \diff W^k_t},
\end{equation}
where $a$ is the Lagrangian label; some transport models based on a hierarchy of Markov models have been
studied in a series of works~\cite{BM2002,BM2003}.  
The assumption \eqref{eq:barx} leads to, for example, the stochastic QG equation
\begin{equation}
  \diff q_i({\bf x},t) + (\bar{\bf u}_i({\bf x},t)\diff t + {\color{red}\sum\limits^K_{k=1}\xi^{k}_i({\bf x})\circ
  \diff W^k_t})\cdot\nabla q_i({\bf x},t) = (F_i-\beta\partial_x\psi_i({\bf x},t))\diff t,
\end{equation}
where $i=1,2$, and $F_i$ being the right hand side of~\eqref{eq:pv}.
This is the system of stochastic PDEs that we actually solve. That is, equation \eqref{eq:barx} is not
explicitly solved but describes the motion of fluid particles under
the stochastic PDE solution.

The goal of the stochastic PDE is to model the coarse-grained
components of a deterministic PDE that exhibits rapidly fluctuating
components. The derivation of deterministic fluid dynamics starts
from the equation
\begin{equation}
  \label{eq:dx}
\diff {\bf x}(a,t) = {\bf u}({\bf x}(a,t),t)\diff t.
\end{equation}
After defining an averaged trajectory $\bar{\bf x}(a,t)$, we write
\begin{equation}
  {\bf x}(a,t) = \bar{\bf x}(a,t) + \eta(\bar{\bf x}(a,t),t/\epsilon^2)
\end{equation}
on the assumption that the fluctuations in $\eta$ in  are faster
than those in $\bar{\bf x}$; this scale separation is parameterised
by the small parameter $\epsilon$. Thus the deterministic equation for
$\bar{\bf x}$ is
\begin{equation}
  \label{eq:dxbar}
  \diff \bar{\bf x}(a,t) = {\bf u}(\bar{\bf x}(a,t) + \eta(\bar{\bf x}(a,t)),t/\epsilon^2)\diff t
  - \eta_t(\bar{\bf x}(a,t),t/\epsilon^2)\diff t.
\end{equation}
If $\eta$ (denoted as $\zeta$ in~\cite{CoGoHo2017})  has a fast dependency on $t$ and has a stationary invariant
measure, then according to homogenisation theory~\cite{CoGoHo2017}
we may average this
equation over the invariant measure (subject to a centring condition)
to get an effective equation
\begin{equation}
  \diff \bar{\bf x}(a,t) = \bar{\bf u}(\bar{\bf x}(a,t), t)\diff t + {\color{red}\sum\limits^K_{k=1}
  \xi^k(\bar{\bf x}(a,t))\circ \diff W^k_t} + \mathcal{O}(\epsilon).
\end{equation}
After truncation of this sum, we recover equation \eqref{eq:barx}.

We assume that ${\bf u}({\bf x},t)$ can be modelled well with a fine grid simulation,
whilst $\bar{\bf u}(\bar{\bf x},t)$ can be modelled with a coarse grid simulation. Then,
we wish to estimate $\xi$'s using data from ${\bf u}({\bf x},t)$ in order to simulate
$\bar{\bf u}(\bar{\bf x},t)$. 

Our methodology is as follows. We spin up a fine grid solution from $t=-T_{spin}$
to $t=0$ (till some statistical equilibrium is reached), then record
velocity time series from $t=0$ to $t=M\Delta t$, where $\Delta
t=k\delta t$, and $\delta t$ is the fine grid timestep. We define
${\bf x}_{ij}^0$ as coarse grid points, 
where $i=1,\ldots,N_y$, $j=1,\ldots,N_x$.

For each $m=0,1,\ldots,M-1$, we
\begin{enumerate}
\item Solve $\dot{\bf x}_{ij}(t)={\bf u}({\bf x}_{ij}(t),t)$ with initial condition ${\bf x}_{ij}(m\Delta t)={\bf x}^0_{ij}$, where 
${\bf u}({\bf x},t)$ is the solution from the fine grid simulation.
\item Compute $\bar{\bf u}(\bar{\bf x}_{ij}(t),t)$ by spatially averaging ${\bf u}({\bf x},t)$ over the coarse grid cell size around gridpoint $(i,j)$.
\item Compute $\bar{\bf x}_{ij}$ by solving the equation
\begin{equation}
\dot{\bar{{\bf x}}}_{ij}(t) = \bar{{\bf u}}(\bar{\bf x}_{ij}(t),t)
\label{eq:x_bar}
\end{equation}
with the same initial condition.
\item Compute the difference $\Delta {\bf x}_{ij}^m = \bar{{\bf x}}_{ij}((m+1)\Delta t) - {\bf x}_{ij}((m+1)\Delta t)$, 
which measures the error between the fine and coarse trajectory.
\end{enumerate}

Having obtained $\Delta {\bf x}_{ij}^m$, we extract the basis
for the noise. This amounts to a Gaussian model of the form
\begin{equation}
 \frac{\Delta {\bf x}_{ij}^m}{\sqrt{\Delta t}} = \Delta \bar{{\bf x}}_{ij} + {\color{red}\sum\limits^K_{k=1} {\xi}^k_{ij}\Delta W^k_m},
\end{equation}
where $\Delta W^k_m$ are independent and identically distributed normal random variables with mean zero and variance one. 

We estimate $\xi$ by minimising
\begin{equation}
\mathbb{E}\left[\left(\sum\limits_{i,j,m}\frac{\Delta {\bf x}_{ij}^m}{\sqrt{\delta t}} - \Delta \bar{{\bf x}}_{ij} - 
{\color{red}\sum\limits^K_{k=1} {\xi}^k_{ij}\Delta W^k_m}\right)^2\right],
\end{equation}
where the choice of $K$ can be informed by using EOFs. 

In order to compute the EOFs, we first compute the matrix of observables 
\begin{equation}
\mathbf{F}:=
\begin{pmatrix}
 \Delta {\bf x}^0_{11} & \Delta {\bf x}^0_{12} & \ldots  & \Delta {\bf x}^0_{N_y,N_x}\\
 \Delta {\bf x}^1_{11} & \Delta {\bf x}^1_{12} & \ldots  & \Delta {\bf x}^1_{N_y,N_x}\\
 \vdots                & \vdots                &  \ddots & \vdots \\
 \Delta {\bf x}^{M-1}_{11} & \Delta {\bf x}^{M-1}_{12} & \ldots  & \Delta {\bf x}^{M-1}_{N_y,N_x}\\
\end{pmatrix},
\end{equation}
and the covariance matrix $\mathbf{R}:=\mathbf{F}^{\rm T}\mathbf{F}/(M-1)$. Then, we compute the eigenvectors of $\mathbf{R}$ (EOFs), 
and use them as $\xi$'s in the stochastic QG model~\eqref{eq:SLTpv}.

Our choice of Empirical Orthogonal Function analysis is based on the capability of this method to extract spatially coherent, temporally uncorrelated and statistically significant modes of transient variability from multivariable time series.
In particular, this method is efficient for dimensionality reduction, compression and spatio-temporal variability analysis 
of atmospheric and oceanic data. Generally speaking, one can use different flow decomposition methods instead of EOF analysis
(e.g. Dynamic Mode Decomposition (DMD)~\cite{Schmid2010}, Optimized DMD~\cite{Chen_et_al2012}, 
Singular Spectrum Analysis~\cite{ElsnerTsonis1996}, etc.), and analyse how they affect the parameterisation, but such a study would be beyond the scope of this paper.

\subsubsection{Approximation of the Lagrangian evolution\label{sec:approx_lagrangian_evol}}
In this section, we apply EOF analysis to the fluctuating component of 
$\Delta {\bf x}$ and perform uncertainty
quantification tests for the stochastic differential equation (SDE)~\eqref{eq:barx} by comparing the true deterministic solution with the ensemble of stochastic equations.
As the true deterministic solution, we take the solution $\bar{\bf x}^c:=\bar{\bf x}(t)$ of the deterministic equation~\eqref{eq:x_bar}.
The stochastic ensemble is given by the solution $\overline{\bf x}(t)$
of SDE~\eqref{eq:barx} computed for independent realizations of the Brownian noise $W$; 
Lagrangian particles are remapped to their original positions (the nodes of the Eulerian grid $G^c=129\times65$) every time step $\Delta t$. 

In order to solve the deterministic equation~\eqref{eq:x_bar}, we use the classical 4-stage 
Runge--Kutta method~\cite{HNW1993}. 
The SDE~\eqref{eq:barx} is solved with the stochastic version of the Runge--Kutta method presented by Algorithm~\ref{alg:SRK4}.

\begin{algorithm}
\caption{Stochastic Runge--Kutta method}
\label{alg:SRK4}
\begin{algorithmic}
\FOR{$n=0,1,2,\ldots$}
\STATE{
\begin{tabular}{ll}
${\bf k}_1={\bf u}(\bar{\bf x}_n,t_n)$, & ${\bf l}_1={\color{red}\Xi(\bar{\bf x}_n,t_n)}$,\\[0.125cm]
${\bf k}_2={\bf u}(\bar{\bf x}_n+\frac{\Delta t}{2}{\bf k}_1,t_n+\frac{\Delta t}{2})$, & ${\bf l}_2={\color{red}\Xi(\bar{\bf x}_n+\frac{\Delta W}{2}{\bf l}_1,t_n+\frac{\Delta W}{2})}$,\\[0.125cm]
${\bf k}_3={\bf u}(\bar{\bf x}_n+\frac{\Delta t}{2}{\bf k}_2,t_n+\frac{\Delta t}{2})$, & ${\bf l}_3={\color{red}\Xi(\bar{\bf x}_n+\frac{\Delta W}{2}{\bf l}_2,t_n+\frac{\Delta W}{2})}$,\\[0.125cm]
${\bf k}_4={\bf u}(\bar{\bf x}_n+\Delta t{\bf k}_3,t_n+\Delta t)$, & ${\bf l}_4={\color{red}\Xi(\bar{\bf x}_n+\Delta W{\bf l}_3,t_n+\Delta W)}$,\\
\end{tabular}}
\hspace*{-1cm}
\STATE{
\begin{equation}
\bar{\bf x}_{n+1}=\bar{\bf x}_n+({\bf k}_1+2({\bf k}_2+{\bf k}_3)+{\bf k}_4)\frac{\Delta t}{6}+{\color{red}({\bf l}_1+2({\bf l}_2+{\bf l}_3)+{\bf l}_4)\frac{\Delta W}{6}}.
\label{eq:srk4}
\end{equation}}
\ENDFOR
\end{algorithmic}
\end{algorithm}
\noindent
Here $\{\bar{\bf x}_i\}^{N_L}_{i=1}$ is the vector of coordinates of Lagrangian particles, 
with $N_L=N_xN_y$, being the number of Lagrangian particles; in this case $N_x=129$, $N_y=65$. 
The stochastic term $\Xi(\bar{\bf x},t)$ is given by
\begin{equation}
\Xi(\bar{\bf x},t):={\color{red}\sum\limits^K_{k=1}\xi^k(\bar{\bf x}(t)) \Delta W^k(t)}.
\end{equation}

Before passing to numerical results, we note that the size of the time step is a critical component for uncertainty quantification and 
data assimilation.
The time step should be short enough that the stochastic ensemble encompasses the true solution. 
Our experiments show that $\Delta t=24\, \rm hours$ properly fulfils this condition.
Note that the stochastic terms in the SPDE are introduced to model the difference between the truth/resolved 
solution and the coarse-grained model. This coarse-graining takes place in space and time, 
and so the difference will depend on both the lengthscale and the timescale of the coarse graining process. 
Hence, the choice of time interval to compute the coarse graining will have an influence on the choice of the stochastic term 
(i.e., the $\xi$'s).

\begin{figure}[htp]
\begin{center}
\includegraphics[scale=0.2]{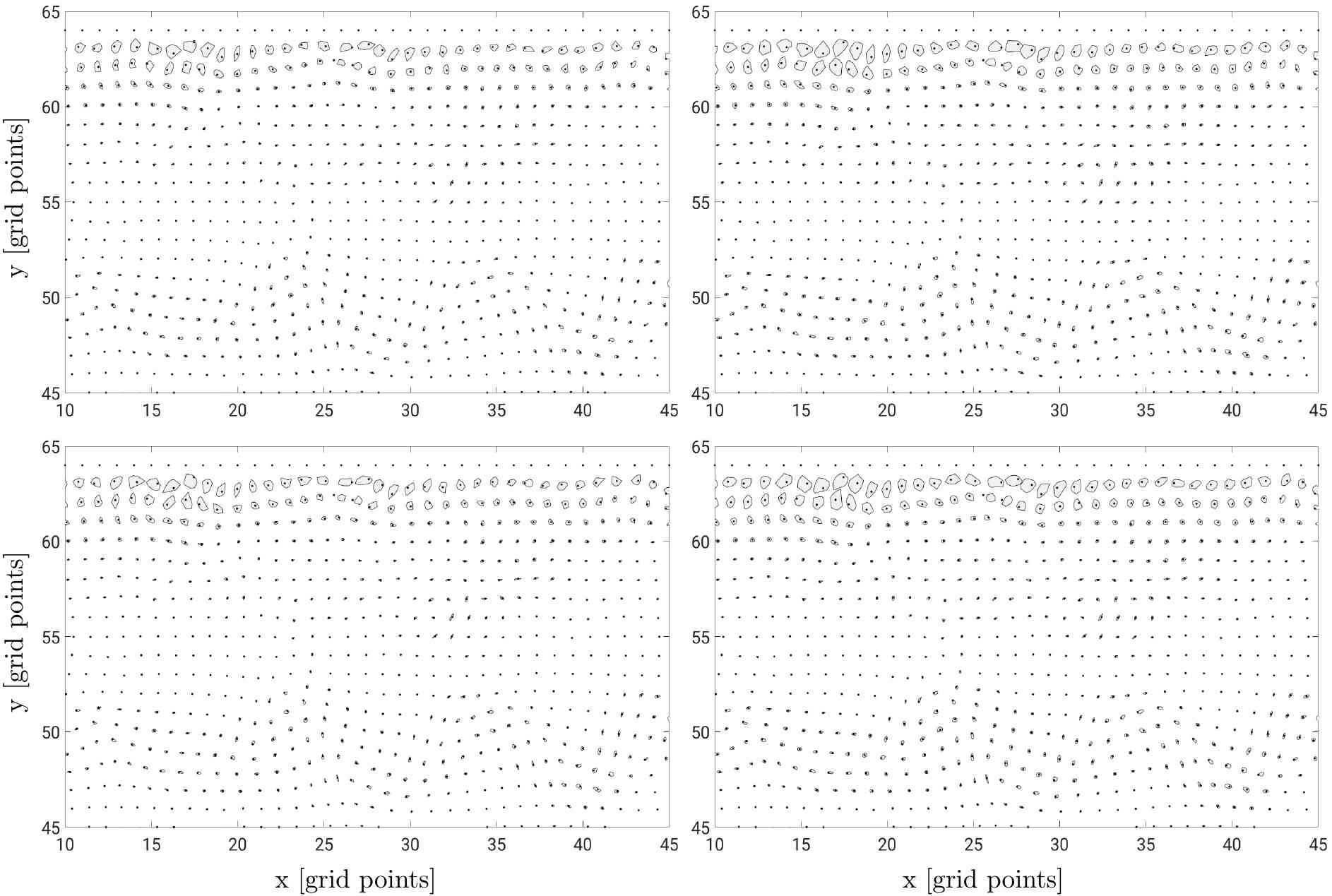}
\caption{Shown is a typical dependence of the area of the stochastic cloud $A^c$ 
on the size of the stochastic ensemble $\overline{\bf x}$. 
The left and right column shows the area of the stochastic cloud (marked in grey color) which consists of $N=100$ and $N=400$ ensemble members, respectively.
The stochastic ensemble has been computed for the \textbf{\textit{first 64 leading EOFs capturing 96\% of the flow variability}} 
(top row)
and the \textbf{\textit{first 128 leading EOFs capturing 99\% of the flow variability}} (bottom row). 
The true solution $\bar{\bf x}^c$ is marked with a black dot.
The plot represents a typical part of the computational domain of size $[10,45]\times[45,65]$ in the first layer, which can be divided into two regions: a fast flow region 
(the boundary layer along the northern boundary $[10,45]\times[60,65]$, the jet occupying the domain $[10,45]\times[45,52]$) and a slow flow region $[10,45]\times(52,60)$.
}
\label{fig:eof64_100_400}
\end{center}
\end{figure}

First, we demonstrate that the true deterministic solution $\bar{\bf x}^c$ is enclosed within a cloud of stochastic solutions $\mathcal{S}$ (also referred as a stochastic spread).
To this end, we study how the area of the stochastic cloud 
(defined as the convex hull of the particle positions and denoted by $A^c$) depends on both the size of the stochastic ensemble, $N$, 
and the number of EOFs denoted by $K$. The results are presented in Figure~\ref{fig:eof64_100_400}.

We identify three key parameters which influence the area of the stochastic cloud: the number of EOFs,
the size of the stochastic ensemble, and the flow velocity.
As Figure~\ref{fig:eof64_100_400} shows, 
the more EOFs are used in the stochastic model, the wider the stochastic spread becomes. 
This difference is more pronounced in the fast flow region, although it is relatively small, since the
variance captured by 64 and 128 EOFs differs only by 3\%.
The same is true for the size of the stochastic ensemble and the velocity of the flow. Namely, the stochastic cloud widens 
as the ensemble size or the flow velocity increases.
The velocity of the flow contributes much more to the size of the spread than the number of ensemble members or EOFs
(compare the area of the stochastic cloud in the fast and slow region for different number of ensemble members). 
The results show that regardless of the flow dynamics the true solution $\bar{\bf x}^c$ lies within the stochastic cloud almost everywhere. 

\begin{figure}[htp]
\begin{center}
\includegraphics[scale=0.2]{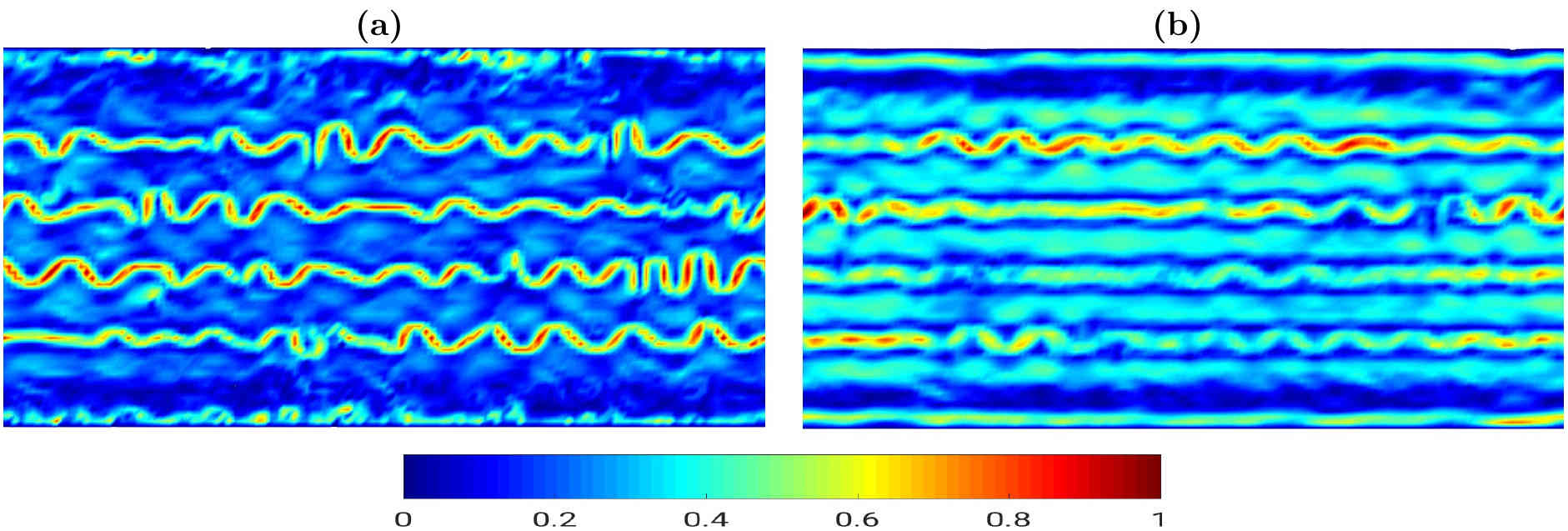}
\caption{Shown are {\bf (a)} instantaneous and {\bf (b)} time-averaged normalized velocity fields for the heterogeneous flow.
}
\label{fig:vel_fields}
\end{center}
\end{figure}

To see a more global picture, we divide the flow dynamics into fast 
(the northern and southern boundary layers, and the jets) and 
slow (flow between the jets), and show instantaneous and time-averaged normalized velocity fields 
in Figures~\ref{fig:vel_fields}a and~\ref{fig:vel_fields}b, respectively.
In terms of the flow velocity, we quantify the flow by the Reynolds number defined as
\begin{equation}
Re:=\left<u\right> Rd_1/\nu, 
\end{equation}
where $\left<u\right>$ is the maximum time-mean velocity, and $Rd_1$ is the first baroclinic Rossby deformation radius. We remark that $Re$ can be defined by using different
velocity and length scales (e.g.~\cite{SiegelEtAl2001}). Our definition is focused on the mesoscale eddies characterized by the length scale up to $O(100)\, {\rm km}$ and jets.
In terms of the Reynolds number, the flow decomposition is given by $Re_s<432$ and $432\le Re_f\le 1440$, where $Re_s$ and $Re_f$ are the Reynolds numbers for the slow and fast flow dynamics, respectively.

As seen in Figure~\ref{fig:vel_fields}, the fast velocity regions appear intermittently between slow velocity regions. In particular,
the fast flow region includes the jet-like structure and the boundary flows along the northern and southern boundary, while
the slow flow regions mainly comprise the flows between the jets.

\begin{figure}[htp]
\begin{center}
\hspace*{-2.5cm}
\begin{tabular}{cc}
\begin{minipage}{0.02\textwidth}\rotatebox{0}{$\overline{A}^c$}\end{minipage} & \begin{minipage}{0.75\textwidth}\includegraphics[width=12cm]{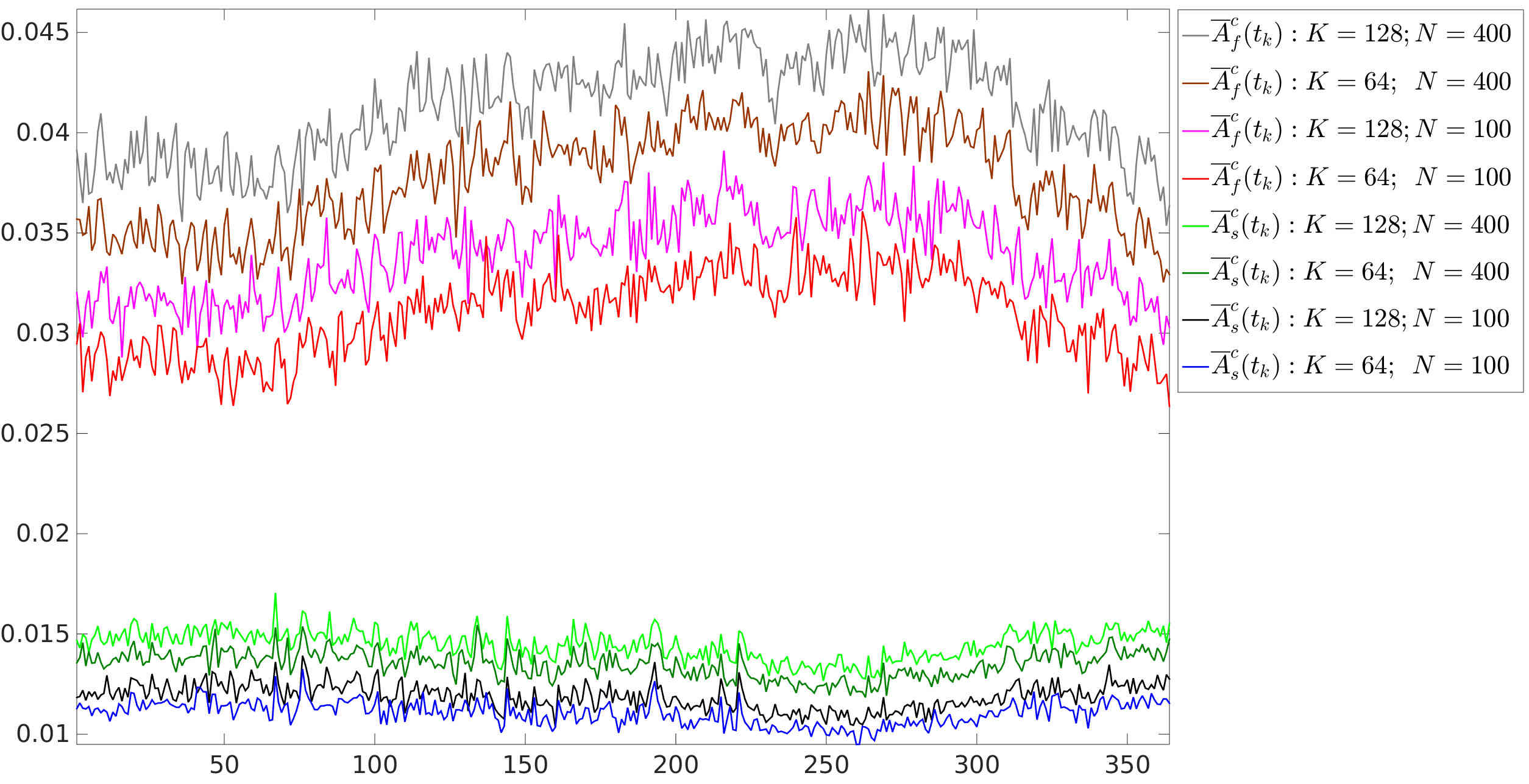}\end{minipage}\\
 & \begin{minipage}{0.5\textwidth}\hspace*{2.5cm} $t\, {\rm [days]}$\end{minipage}\\
\end{tabular}
\caption{Shown is the dependence of the averaged area of the stochastic cloud for the slow, $\overline{A}^c_s$, and fast, $\overline{A}^c_f$, flow regions on
the number of EOFs, $K$, and the size of the stochastic ensemble $N$.
}
\label{fig:cloud_area_fast_slow_flow}
\end{center}
\end{figure}

Before going into detail, it is helpful to introduce the area of the stochastic cloud $\overline{A}^c_s$ and $\overline{A}^c_f$ averaged 
over the number of Lagrangian particles in the slow, $N_{L,s}$, and fast, $N_{L,f}$, flow region, respectively: 
\begin{equation}
\overline{A}^c_s(t_k):=\frac{1}{N_{L,s}}\sum\limits^{N_{L,s}}_{i=1}A^c_{s,i}(t_k),\quad 
\overline{A}^c_f(t_k):=\frac{1}{N_{L,f}}\sum\limits^{N_{L,f}}_{i=1}A^c_{f,i}(t_k),
\end{equation}
where $A^c_{s,i}$ and $A^c_{f,i}$ are the area of the stochastic cloud 
corresponding to the $i$-th true Lagrangian particle 
belonging to the slow and fast flow, respectively.
The dependence of the averaged area of the stochastic cloud for the slow and fast flow regions on the number of EOFs and the size of the stochastic ensemble is presented 
in Figure~\ref{fig:cloud_area_fast_slow_flow}.

Upon analysing the results presented in Figure~\ref{fig:cloud_area_fast_slow_flow},
we have found that they are in good agreement with the analysis of instantaneous snapshots
(Figure~\ref{fig:eof64_100_400}). In particular, the averaged area of the stochastic cloud $\overline{A}^c$ is influenced by the 
three parameters we identified for the case of instantaneous flows: the number of EOFs, the size of the stochastic ensemble, and the flow velocity. More importantly, the 
qualitative behavior of $\overline{A}^c$ is similar to those of $A^c$ (the area of the stochastic ensemble associated with a given Lagrangian particle). 
Namely, as the size of the stochastic ensemble increases so does the area of the cloud (for example, compare the red and brown lines for which the ensemble size is 
$N=100$ and $N=400$, respectively). The increase in the number of EOFs also leads to a larger area of the cloud 
(for instance, compare the red and magenta lines for which $K=64$ and $K=128$, respectively). However, it is not a significant increase, since
the variance captured by 96 and 128 EOFs differs only by 3\%.
These results stay the same for the stochastic QG model studied in Section~\ref{sec:xi_SQG}.
The same is true for the flow velocity: the faster the flow, the larger the stochastic cloud (compare the red and blue lines which corresponds to the fast and 
slow velocity region, respectively). The most important observation here is that 
we can estimate the contribution of each parameter to the parameterisation. The size of the stochastic ensemble and the number of EOFs
have a rather small effect on the area of the stochastic cloud. The most significant contribution comes from the velocity of the flow. Namely,
the size of the stochastic cloud for fast flows is always larger than that for slow flows (compare the upper four curves with the lower four curves in Figure~\ref{fig:cloud_area_fast_slow_flow}).

\subsubsection{Initial conditions\label{sec:ICs}}
The choice of the initial condition for the stochastic QG model is important, 
especially in the context of uncertainty quantification and data assimilation, for it significantly influences the 
evolution of the flow as well as its further predictability.
A straightforward approach based on a random perturbation of the true solution at time $t=0.0$
can inject unphysical perturbations into the flow which, in turn, can result in an unphysical solution.
Therefore, in order to perform uncertainty quantification tests
we need a number of independent realizations of the initial condition that 
are physically consistent with the flow dynamics. Namely, each independent realization of the
initial condition is advected by the stochastic QG model until $t_0=0$, and thus, it is a solution to the SPDE.
To this end, we start at time $t=-t^*$ with the true solution $q^a$ of the deterministic model and run it until
$t_0=0$ with independent realizations of the Brownian noise $W$ 
to produce independent samples from the initial condition. 
As a result, the ensemble of stochastic solutions (also referred to as ensemble members or particles), $\mathcal{S}(x,y)$, 
covers the true deterministic solution at time $t_0$.

We initialise the ensemble in a manner consistent with performing ensemble data assimilation (using a particle filter, for example). 
The ensemble of initial data represents our belief about the coarse-grained solution state according to a particular 
scenario, which would be updated by the filter using observational data. 
It is important to note that in all the numerical simulations which follow, the initial stochastic ensemble is
biased (if not stated otherwise), i.e. the ensemble mean is not equal to the true solution at 
time $t=0$, since the ensemble mean can drift away from the true solution over the initial 
spin up interval $T$ (see below) needed for computing the initial condition for the stochastic model.
This is of course typical of a data assimilation scenario: assimilating data could reduce this 
bias at future times. In this work we do not assimilate data, but just investigate whether the ensemble spread is likely to be sufficient 
for stable data assimilation.

The goal of the next experiment is to study for how long this property holds. To this end, we introduce the following function 
\begin{equation}
T_{\mathcal{S}}:=\frac{1}{|T|}\int\limits_{T}\delta(q^a) \, dt,\qquad
\delta(q^a)=\left\{
 \begin{aligned}
  1\quad \text{if } q^a\in\mathcal{S}(x,y),\\
  0\quad \text{if } q^a\not\in\mathcal{S}(x,y),
 \end{aligned}
 \right.
\label{eq:T_S} 
\end{equation}
which represents the time period spent by the true deterministic solution, 
$q^a$, within the spread of stochastic solutions defined for each coordinate $(x,y)$ as
$\mathcal{S}(x,y):=[\min (q^p(x,y,W_t,t)),\max (q^p(x,y,W_t,t))]$, where the
minimum and maximum are computed over the stochastic ensemble,
and $q^p(x,y,W_t,t)$ is the solution of the parameterised QG model.

We carried out two numerical experiments to compute $T_{\mathcal{S}}$.
In the first one, we start at time $t^*=-1$ hour with 
the true solution $q^a(t^*)$ and run the stochastic QG model~\eqref{eq:SLTpv} until $t_0=0$ with 100 independent realizations of the Brownian noise $W$,
and with the first 64 leading EOFs, which capture 96\% of the total variance. In the second one, we do the same, but start at $t^*=-16$ hours.
Our results (not presented here) show
that the spread of stochastic solutions $\mathcal{S}$ 
captures the true deterministic velocity ${\bf u}^a$, stream function $\psi^a$, and PV anomaly $q^a$ for the heterogeneous and homogeneous flows
in both experiments equally well almost everywhere 
in the computational domain except the neighbourhood of the northern and southern boundaries. 
This boundary layer dynamics is difficult to capture on the coarse grid, because of the low resolution.
However, the boundary layer is very small with respect to the whole domain, and so its contribution to uncertainty quantification results is miniscule.
The length of the time interval $T$ has a minor influence on the behavior of the spread 
thus ensuring a better coverage of the true solution with the spread over a longer time period.
Overall, we have shown that the stochastically advected deterministic initial condition provides a solid basis 
for uncertainty quantification tests and data assimilation, which will be the object of future research.

\subsubsection{Uncertainty quantification for the stochastic QG model\label{sec:xi_SQG}}
\begin{figure}[htp]
\begin{center}
\includegraphics[scale=0.125]{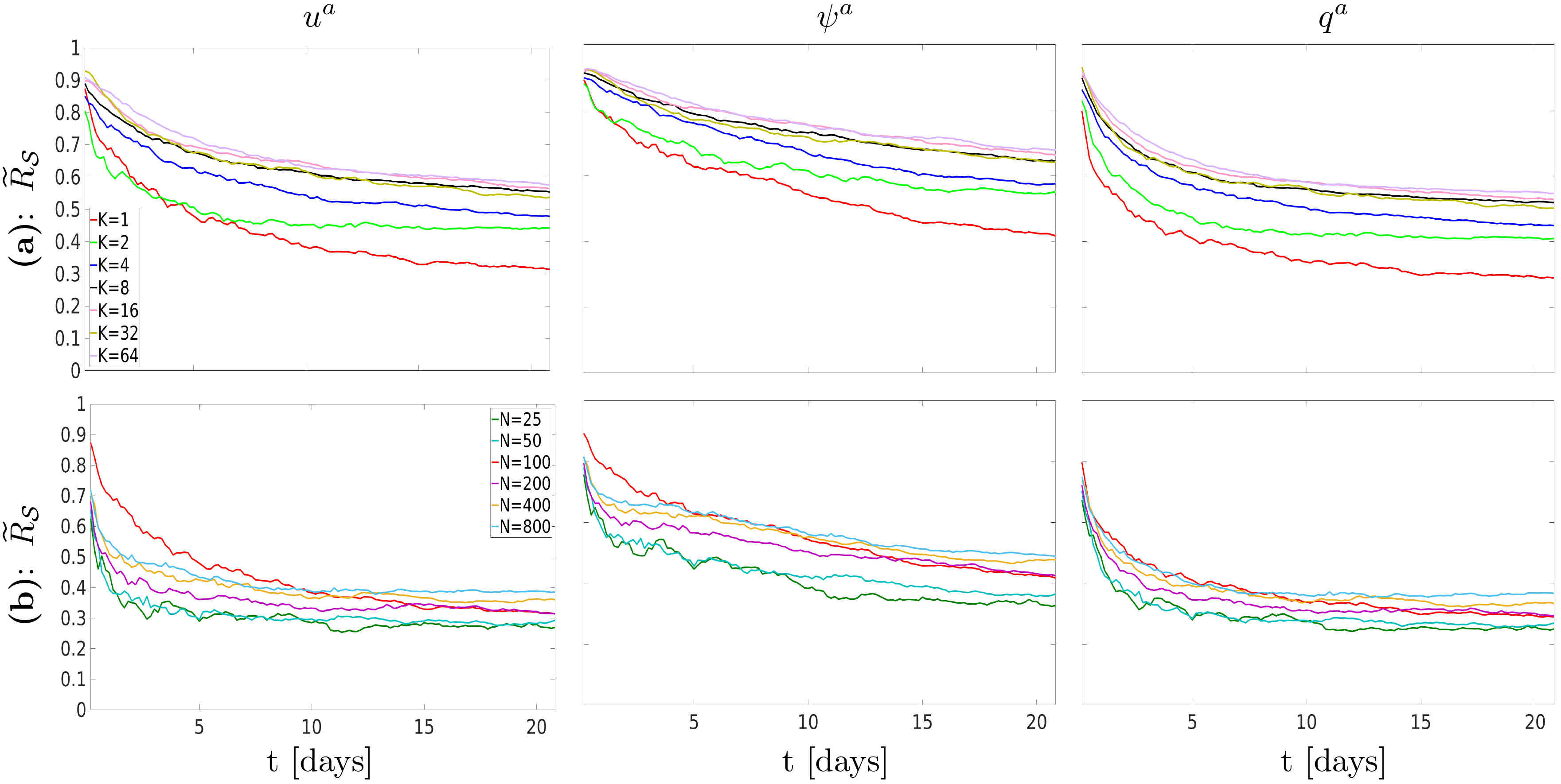}
\caption{
Shown is the dependence of $\widetilde{R}_{\mathcal{S}}$
for the velocity component $u^a$ ($v^a$ is not shown, since it behaves qualitatively similar to $u^a$), stream function $\psi^a$, 
and PV anomaly $q^a$ on {\bf (a)} the number of EOFs $K$ ($N=100$ in this case) 
and {\bf (b)} size of the stochastic ensemble $N$ ($K=1$ in this case) 
over the time period $T=[0,21]$ days for the 
\textbf{\textit{heterogeneous flow}} in Figure~\ref{fig:qf_qa_qc_qam_qcm_mu1D-8_129x65}; $G^c=129\times65$. 
Using $K=\{1,2,4,8,16,32,64\}$ leading EOFs allows to capture 23\%, 42\%, 60\%, 77\%, 89\%, 96\%, and 99\% of the flow variability, respectively.
The initial conditions for the stochastic model have been computed over the spin up period $T_{\rm spin}=[-8,0]$ hours.
}
\label{fig:R_EOF_ensemble_mu4D-8}
\end{center}
\end{figure}

\begin{figure}[htp]
\begin{center}
\includegraphics[scale=0.125]{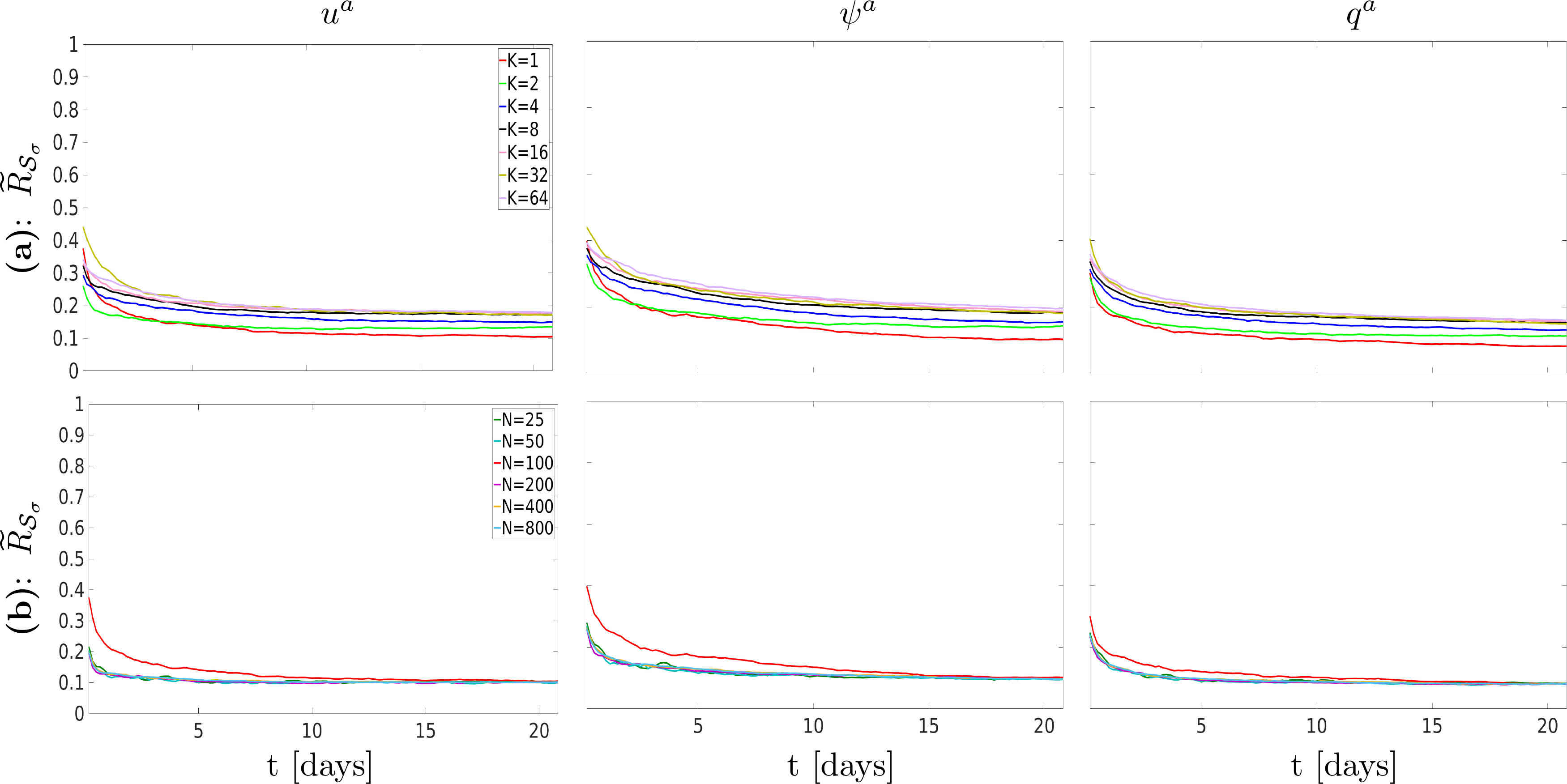}
\caption{The same as in Figure~\ref{fig:R_EOF_ensemble_mu4D-8}, but for $\widetilde{R}_{\mathcal{S}_{\sigma}}$.}
\label{fig:R_EOF_ensemble_mu4D-8_std}
\end{center}
\end{figure}

The Lagrangian evolution studied above demonstrates encouraging results. However, it cannot guarantee that 
the application of EOFs to the stochastic QG equations~\eqref{eq:SLTpv} is equally beneficial. 
Therefore, this section focuses on uncertainty quantification for the stochastic QG model. 
Here, we analyse the behaviour of the stochastic ensemble for two different cases. 
In the first case (Figure~\ref{fig:R_EOF_ensemble_mu4D-8}), we study how many points (each particular point represents the true deterministic solution at this point) 
in the computational domain remain within the spread, i.e. we analyse how the function
\begin{equation}
\widetilde{R}_{\mathcal{S}}(q^a)=\frac{1}{N_x N_y}\sum\limits^{N_x}_{i=1}\sum\limits^{N_y}_{j=1}
\delta(q^a_{ij}),\qquad
\delta(q^a_{ij})=\left\{
 \begin{aligned}
  1\quad \text{if } q^a_{ij}\in\mathcal{S}(x,y),\\
  0\quad \text{if } q^a_{ij}\not\in\mathcal{S}(x,y),
 \end{aligned}
 \right.
\label{eq:R_S} 
\end{equation}
depends on the number of EOFs and the ensemble size.
In the second case (Figure~\ref{fig:R_EOF_ensemble_mu4D-8_std}), we impose more stringent restrictions upon the parameterisation, and 
analyse how many points in the domain remain within one standard deviation of the ensemble mean, 
not within the whole ensemble as before. Namely, we study how the function
\begin{equation}
\widetilde{R}_{\mathcal{S}_{\sigma}}(q^a)=\frac{1}{N_x N_y}\sum\limits^{N_x}_{i=1}\sum\limits^{N_y}_{j=1}
\delta(q^a_{ij}),\qquad
\delta(q^a_{ij})=\left\{
 \begin{aligned}
  1\quad \text{if } q^a_{ij}\in\mathcal{S}_{\sigma}(x,y),\\
  0\quad \text{if } q^a_{ij}\not\in\mathcal{S}_{\sigma}(x,y),
 \end{aligned}
 \right.
\label{eq:R_S2} 
\end{equation}
depends on the number of EOFs and the ensemble size.
Here, $\mathcal{S}_{\sigma}(x,y):=[\bar{q}^p-\sigma(q^p),
\bar{q}^p+\sigma(q^p)]$,
where $\sigma(q^p)$ is the standard deviation of the ensemble mean:
\begin{equation}
\bar{q}^p:=\frac{1}{N}\sum\limits^N_{n=1}q^p_n(x,y,W_t,t).
\end{equation}

As seen in Figures~\ref{fig:R_EOF_ensemble_mu4D-8} and \ref{fig:R_EOF_ensemble_mu4D-8_std}, the smoother the field is, 
the more points in the domain remain within the spread. 
Namely, the stream function $\psi^a$ is enclosed within the spread in more points in the domain compared to the velocity $u^a$ 
(as well as $v^a$, not shown) and PV anomaly $q^a$.
Using more EOFs leads to a better coverage of the truth. For example, using $K=\{1,2,4\}$ EOFs gives 
a spread which gradually captures more individual points in the domain.
However, this effect does not last for long. In particular, for $K>4$ one does not observe a significantly better coverage compared
with $K=4$.
The same conclusion holds for the size of the stochastic ensemble:
the larger the size of the ensemble, the more points in the domain
remain covered with the spread. However, the variation in the ensemble size has a much more minor effect on the spread
than the variation in the number of EOFs.
Obviously, the ratio of individual points enclosed into one standard deviation of the mean of the stochastic spread is much
less than that contained in the full spread (compare Figures~\ref{fig:R_EOF_ensemble_mu4D-8} and~\ref{fig:R_EOF_ensemble_mu4D-8_std}).

\begin{figure}[htp]
\begin{center}
\includegraphics[scale=0.135]{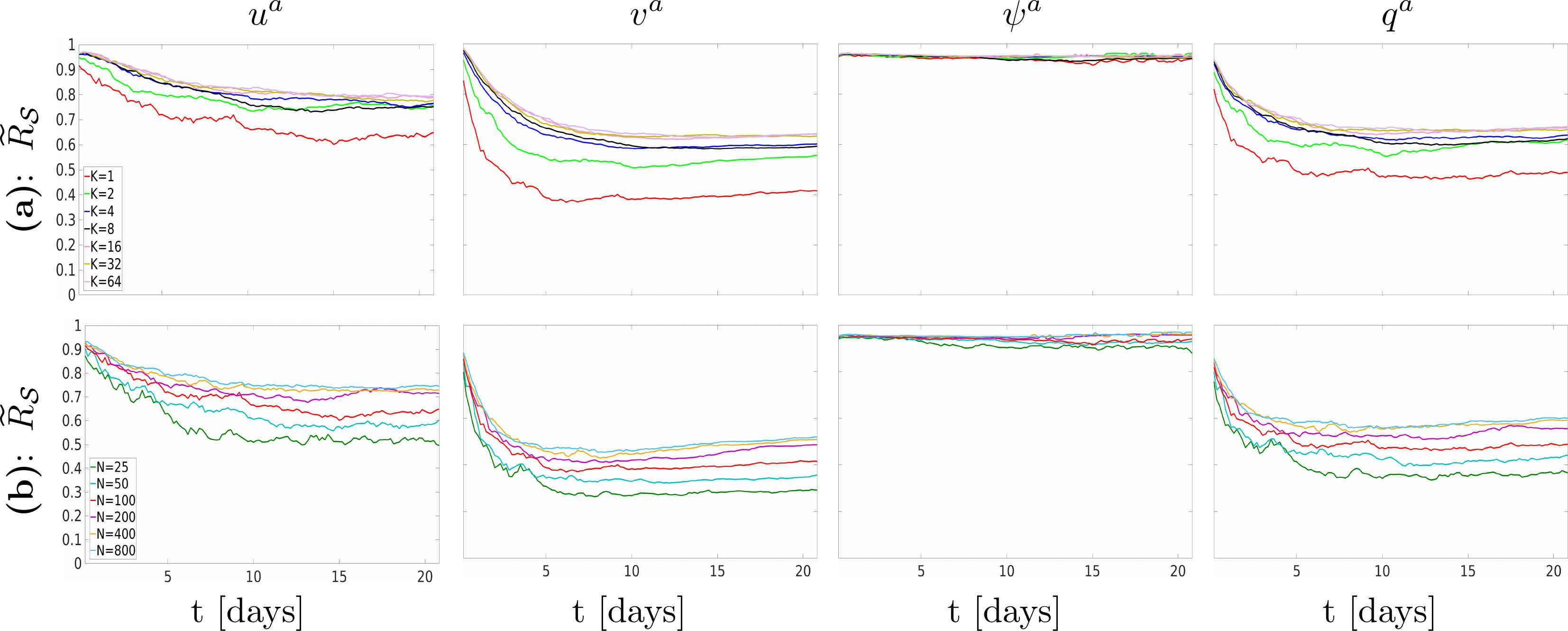}
\caption{The same as in Figure~\ref{fig:R_EOF_ensemble_mu4D-8}, but for $G^c=257\times129$.
We show $v^a$ component of the velocity field, since it behaves differently from $u^a$.
For the higher resolution,
using $K=\{1,2,4,8,16,32,64\}$ leading EOFs allows to capture 39\%, 61\%, 77\%, 90\%, 97\%, 99\%, and 99.8\% of the flow variability, respectively.
}
\label{fig:R_EOF_ensemble_mu4D-8_257x129}
\end{center}
\end{figure}

\begin{figure}[htp]
\begin{center}
\includegraphics[scale=0.135]{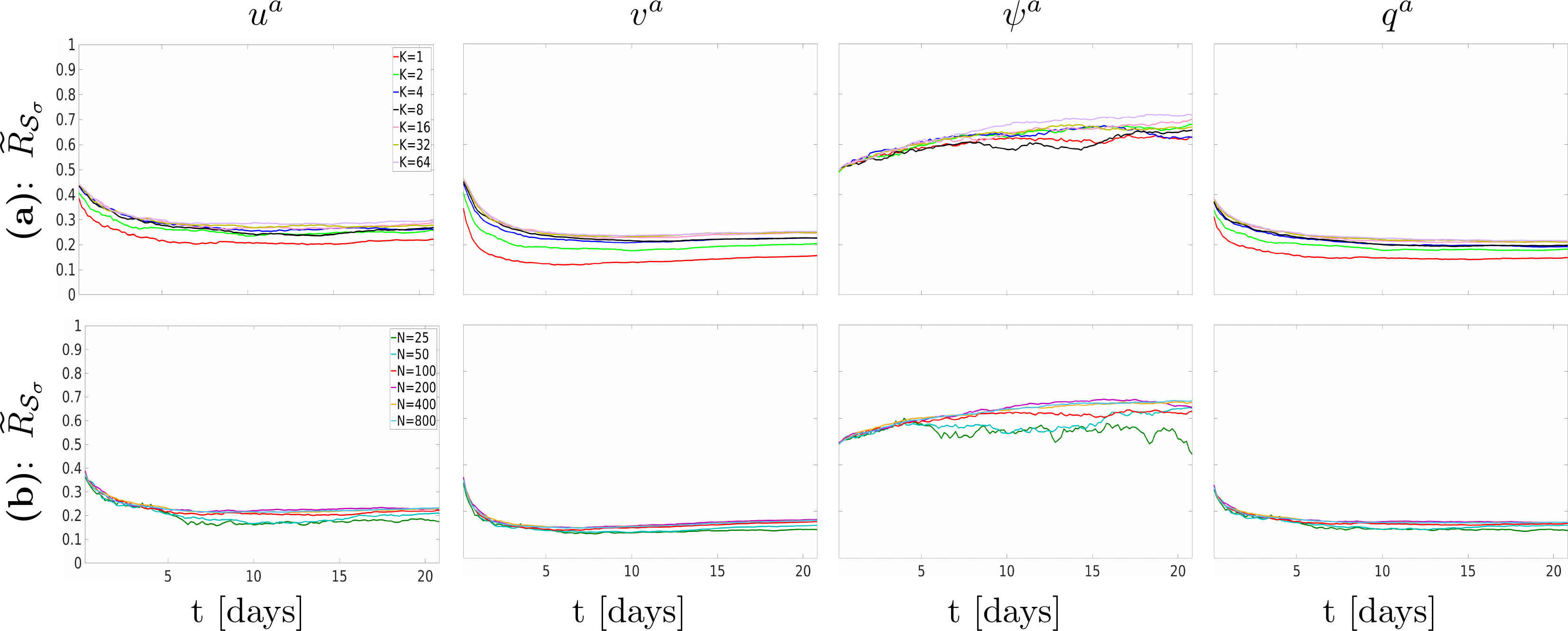}
\caption{The same as in Figure~\ref{fig:R_EOF_ensemble_mu4D-8_257x129}, but for $\widetilde{R}_{\mathcal{S}_{\sigma}}$.}
\label{fig:R_EOF_ensemble_mu4D-8_std_257x129}
\end{center}
\end{figure}

\ansA{Next we study how the resolution influences the effect of the parameterisation, computing} $\widetilde{R}_{\mathcal{S}}$ and $\widetilde{R}_{\mathcal{S}_{\sigma}}$ 
on a grid $G^c=257\times129$ (Figures~\ref{fig:R_EOF_ensemble_mu4D-8_257x129} and~\ref{fig:R_EOF_ensemble_mu4D-8_std_257x129}).
\ansA{As can be seen in Figures~\ref{fig:R_EOF_ensemble_mu4D-8_257x129} and~\ref{fig:R_EOF_ensemble_mu4D-8_std_257x129}, 
many more points in the domain remain covered with
the spread compared with the lower resolution.} Surprisingly, \ansA{the number of covered points even tends to increase in time. This unexpected result is analysed in Section~\ref{sec:deter_vs_stoch}.}
The results for the homogeneous flow \ansA{(not shown)} 
are qualitatively similar to those of the heterogeneous one.

\begin{figure}[htp]
\begin{center}
\includegraphics[scale=0.135]{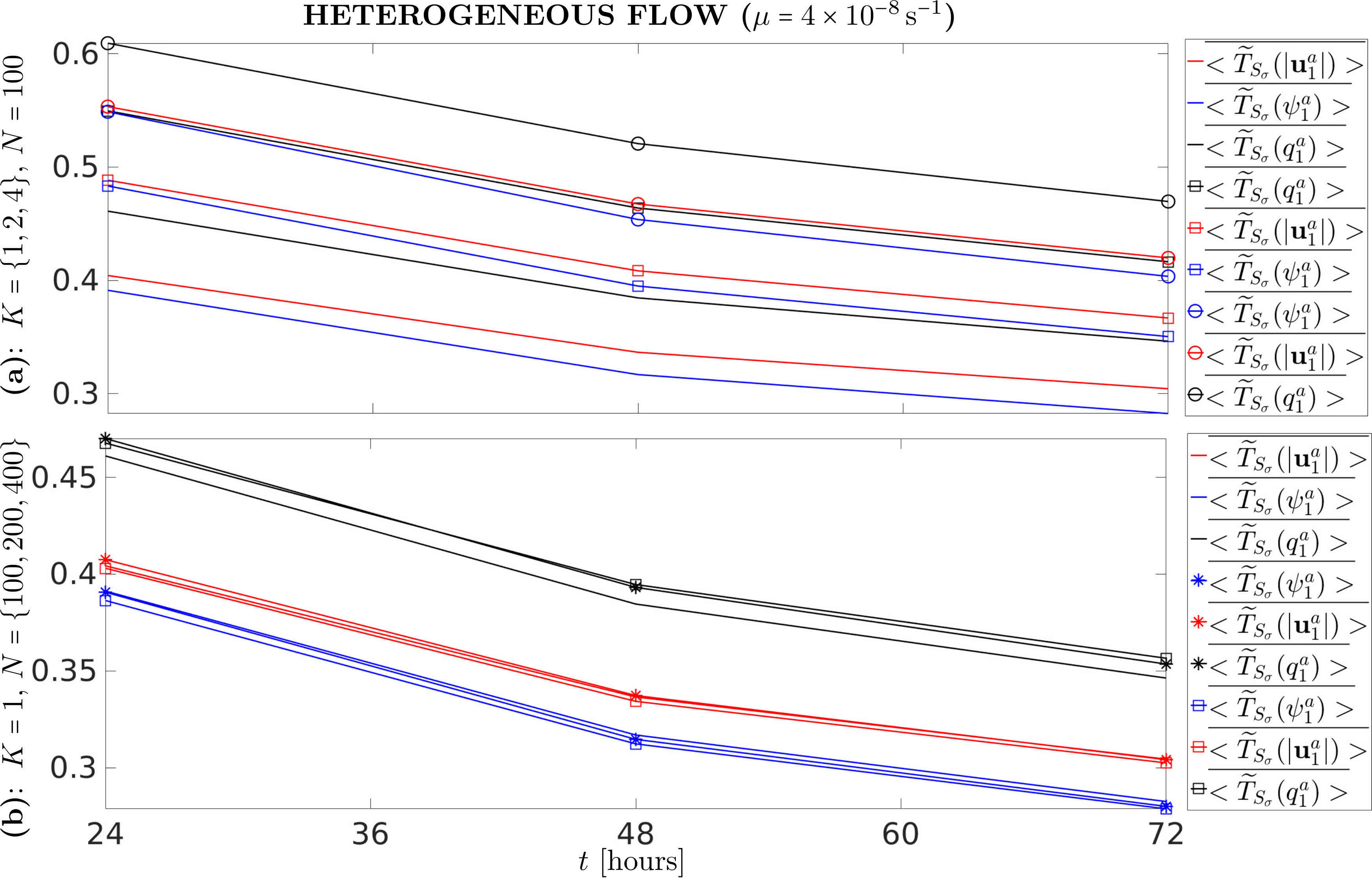}
\caption{Evolution of $\overline{<\widetilde{T}_{\mathcal{S}_{\sigma}}>}$ for 
the heterogeneous flow 
and for different number of EOFs: $K=1$ (solid line), $K=2$ (solid line marked by a cross),
$K=4$ (solid line marked by an circle), as well as for different sizes of the stochastic ensemble: 
$N=100$ (solid line), $N=200$ (solid line marked by an asterisk), $N=400$ (solid line marked by a square).}
\label{fig:std_TimeCoveredBySpread_xi1_2_4_100_200_400particles_mu4D-8_mu4D-7}
\end{center}
\end{figure}

\ansA{To analyse a more global picture, we introduce the function}
\begin{equation}
\widetilde{T}_{\mathcal{S}_{\sigma}}:=\frac{1}{|T|}\int\limits_{T}\delta(q^a) \, dt,\qquad
\delta(q^a)=\left\{
 \begin{aligned}
  1\quad \text{if } q^a\in\mathcal{S}_{\sigma}(x,y),\\
  0\quad \text{if } q^a\not\in\mathcal{S}_{\sigma}(x,y),
 \end{aligned}
 \right.
\label{eq:T_S21} 
\end{equation}
which shows how long the true deterministic solution remains within one standard deviation of the ensemble mean.
\ansA{We examine how $\widetilde{T}_{\mathcal{S}_{\sigma}}$ evolves in time 
at different realizations of initial conditions and the ensemble size.}
We also introduce another function
\begin{equation}
\overline{<\widetilde{T}_{\mathcal{S}_{\sigma}}>}=
\frac{1}{M}\sum\limits^M_{i=1}\frac{1}{|\Sigma|}\int\limits_{\Sigma}\widetilde{T}^{(i)}_{\mathcal{S}_{\sigma}}\, d\Sigma\,,
\end{equation}
which 
is computed with the stochastic QG model started from an initial condition $t_i\,, i=1,2,\ldots,M$.
The function 
$\overline{<\widetilde{T}_{\mathcal{S}_{\sigma}}>}$ is $\widetilde{T}_{\mathcal{S}_{\sigma}}$ averaged
over the computational domain $\Sigma=[0,L_x]\times[0,L_y]$ and over the number of different initial conditions 
(we take $M=10$ in this case, for being limited in computational resources), which are
randomly chosen within the interval of one year (see Figure~\ref{fig:std_TimeCoveredBySpread_xi1_2_4_100_200_400particles_mu4D-8_mu4D-7}). 

The results presented in Figure~\ref{fig:std_TimeCoveredBySpread_xi1_2_4_100_200_400particles_mu4D-8_mu4D-7} show that
the more EOFs is used in the parameterisation the longer the stochastic ensemble can \ansA{cover} the true solution
for both the heterogeneous (Figure~\ref{fig:std_TimeCoveredBySpread_xi1_2_4_100_200_400particles_mu4D-8_mu4D-7}a)
and homogeneous (not shown due to similar behaviour) 
flows.
The same conclusion is true for the ensemble size: 
the larger the size of the ensemble, the longer the spread captures the true solution 
(Figure~\ref{fig:std_TimeCoveredBySpread_xi1_2_4_100_200_400particles_mu4D-8_mu4D-7}b).
However, using yet more ensemble members does not yield a much better coverage of the true solution with the stochastic spread,
especially for the homogeneous flow. 
Another observation is that the stochastic spread gradually loses the track of true solution in time. 

\begin{figure}[htp]
\begin{center}
\includegraphics[scale=0.135]{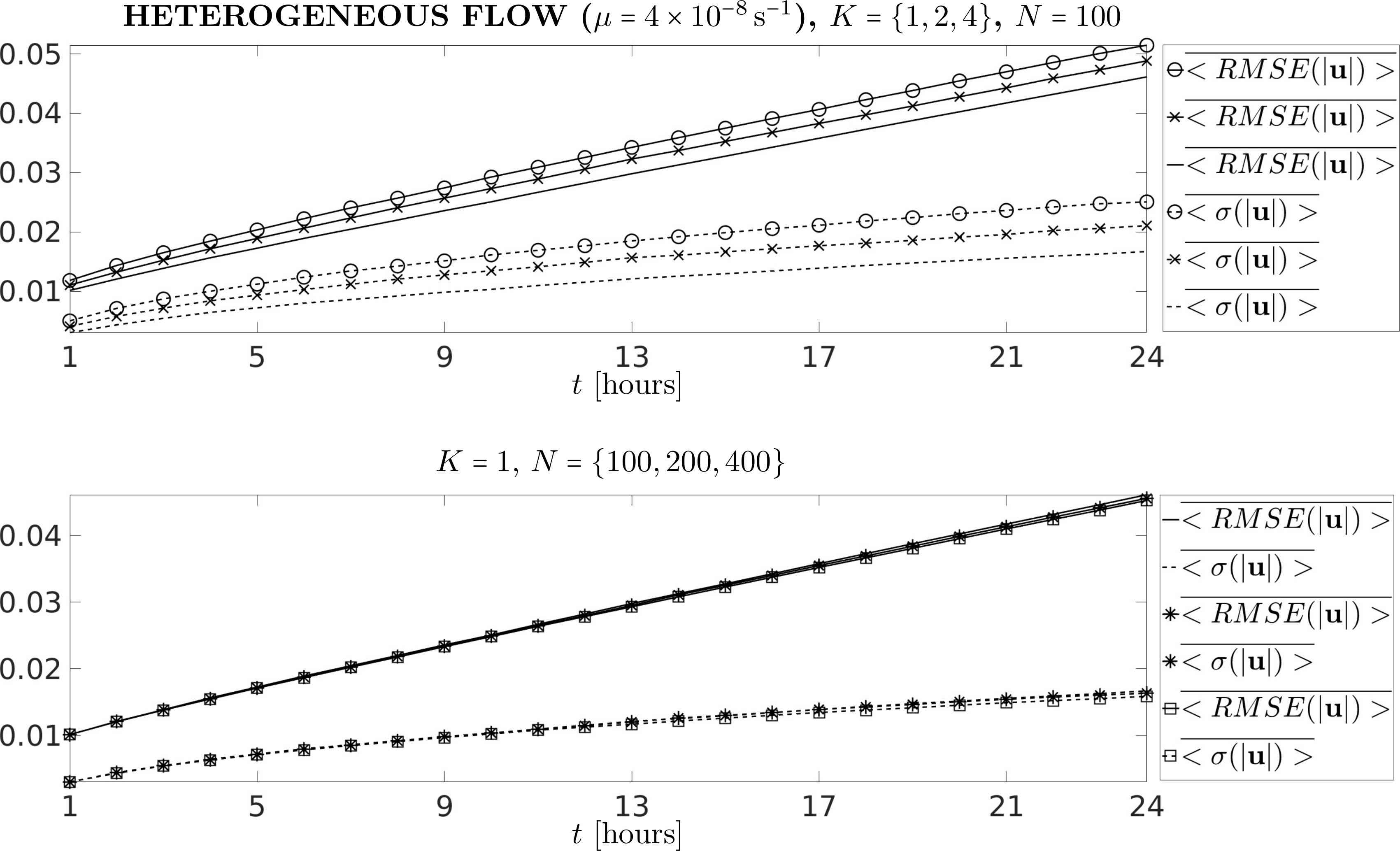}
\caption{Evolution of the root mean square error $\overline{<RMSE(|\boldsymbol{u}|)>}$ (solid line)
and standard deviation $\overline{<\sigma(|\boldsymbol{u}|)>}$ (dashed line) of the ensemble mean for 
the \textbf{\textit{heterogeneous flow}} and different number of EOFs: $K=1$ (solid/dashed line), $K=2$ (solid/dashed line marked by a cross),
$K=4$ (solid/dashed line marked by an circle); and for different sizes of ensemble: 
$N=100$ (solid/dashed line), $N=200$ (solid/dashed line marked by an asterisk), $N=400$ (solid/dashed line marked by a square).
\textit{Note that the initial ensemble is biased.}}
\label{fig:rmse_std_xi1_2_4_100_200_400particles_mu4D-8}
\end{center}
\end{figure}

For a deeper analysis of the stochastic spread, we study its reliability, which is one of the most important
characteristics of the ensemble used in ensemble forecasting (see, e.g.~\cite{Weigel2012}). One of the criteria for ensemble
reliability is that the root mean square error of the ensemble mean, $\overline{<RMSE(|{\bf u}|)>}$, is 
relatively close to the standard deviation of the 
ensemble, $\overline{<\sigma(|\boldsymbol{u}|)>}$. \ansA{In numerical weather prediction, his is called the skill-spread relationship.}
\ansA{Here,} we firstly average these quantities
over the computational domain and different initial conditions $t_i$ $(i=1,2,\ldots,M)$,
and then compute them relative to the their spatial averages which are
denoted as $\overline{<RMSE(|\boldsymbol{u}|)>}$ and $\overline{<\sigma(|\boldsymbol{u}|)>}$, respectively. 
The dependence of
$\overline{<RMSE(|\boldsymbol{u}|)>}$ and $\overline{<\sigma(|\boldsymbol{u}|)>}$ on the number of leading EOFs and the size of the ensemble
over the time period of 24 hours is presented in Figure~\ref{fig:rmse_std_xi1_2_4_100_200_400particles_mu4D-8}. 
\ansA{We can see that there is a mismatch in the spread-skill, which we have explained and corrected in the sequel paper~\cite{CCHWS2020_1}.}
\ansA{We do not show the results for the homogeneous flow, since they are very similar to those of the heterogeneous one.}

We interpret the above results as follows. The difference between the solution of the SPDE approximated on the coarse 
scale and the projection on the same scale of the PDE solution is estimated by drawing  independent samples from the solution 
of the SPDE and comparing their average with the coarsened solution of the PDE. The error between the two is measured 
by the corresponding mean square error $MSE(|\boldsymbol{u}|)$. In the graphs presented in 
Figure \ref{fig:rmse_std_xi1_2_4_100_200_400particles_mu4D-8}, we can see the evolution of the square root of this quantity 
(more precisely its estimate obtained by using 100/200/400 independent samples) denoted by  
$\overline{<RMSE(|\boldsymbol{u}|)>}$. The mean square error  $MSE(|\boldsymbol{u}|)$ is the sum between the variance of the solution of the SPDE and 
the square of the difference between the expected value of the solution of the SPDE and the coarsened solution of the PDE 
(the latter quantity is, in fact, the square of the bias of the approximation). Therefore, the root mean square error  
$\overline{<RMSE(|\boldsymbol{u}|)>}$ is always going to be larger than the standard deviation of the solution of the SPDE,  denoted by   
$\overline{<\sigma(|\boldsymbol{u}|)>}$ . They coincide only when the bias is null. 
As seen in Figure~\ref{fig:rmse_std_xi1_2_4_100_200_400particles_mu4D-8}, 
the difference between the root mean square error $\overline{<RMSE(|\boldsymbol{u}|)>}$ and standard deviation 
$\overline{<\sigma(|\boldsymbol{u}|)>}$, approximated by the square root of the ensemble variance, 
for the heterogeneous and homogeneous flows appears at the very beginning of the simulations thus showing that
the initial stochastic ensemble is biased as expected. This difference is small (1-2\%) and grows slowly in time, 
showing that the bias remains small over the chosen timescale. The number of leading
EOFs and the size of the ensemble members have only a minor effect on both the RMSE and the standard deviation.

\begin{figure}[htp]
\begin{center}
\includegraphics[scale=0.135]{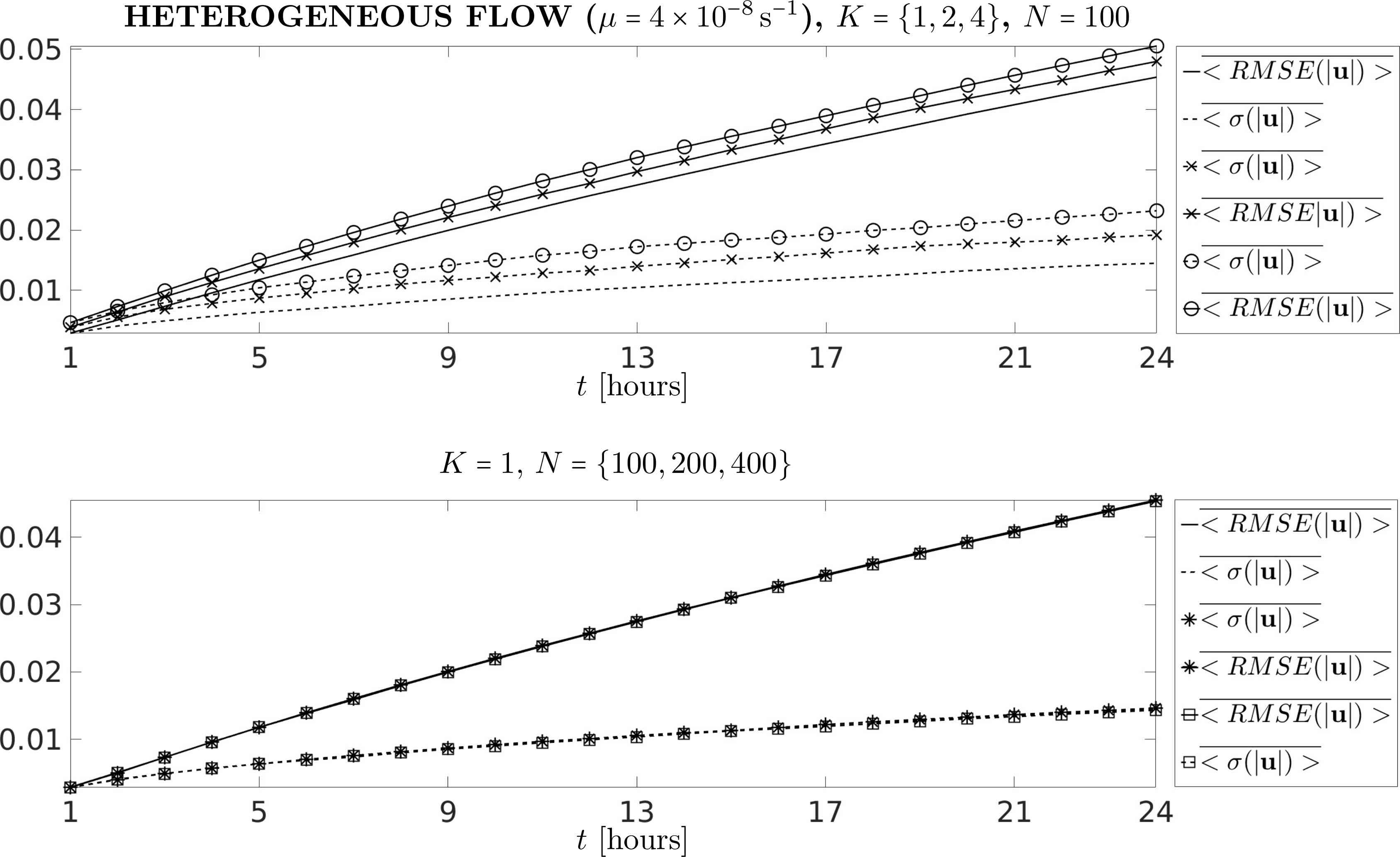}
\caption{The same as in Figure~\ref{fig:rmse_std_xi1_2_4_100_200_400particles_mu4D-8}, but for an unbiased ensemble. 
\ansA{The results for the homogeneous flow look very similar.}}
\label{fig:rmse_std_xi1_2_4_100_200_400particles_mu4D-8_new}
\end{center}
\end{figure}

The bias is reduced when the initial ensemble is unbiased. Of course, in this  
case, $\overline{<RMSE(|\boldsymbol{u}|)>}=\overline{<\sigma(|\boldsymbol{u}|)>}$ at $t=0$.
The evolution of the ensemble unbiased at $t=0$ is shown in 
Figure~\ref{fig:rmse_std_xi1_2_4_100_200_400particles_mu4D-8_new}. 
As for the case of the biased ensemble, using more EOFs as well as the number of ensemble members leads to minor changes in 
$\overline{<RMSE(|\boldsymbol{u}|)>}$ and $\overline{<\sigma(|\boldsymbol{u}|)>}$

Despite the fact the initial ensemble is unbiased, 
the stochastic model still acquires a bias over time, and the
discrepancy between $\overline{<RMSE(|\boldsymbol{u}|)>}$ and $\overline{<\sigma(|\boldsymbol{u}|)>}$ grows in time. 
The bias gets smaller as the grid cell decreases. In the limit in which there is no difference between the grid 
scale on which the PDE is computed and the grid scale on which the SPDE is computed the bias will be zero. 
\ansB{The goal of the parameterisation is not to achieve a zero bias, but to provide a sufficiently large spread that the bias can be reduced by data assimilation, which has been studied in the sequel paper~\cite{CCHWS2020_1}.}      

\begin{figure}[htp]
\begin{center}
\includegraphics[scale=0.135]{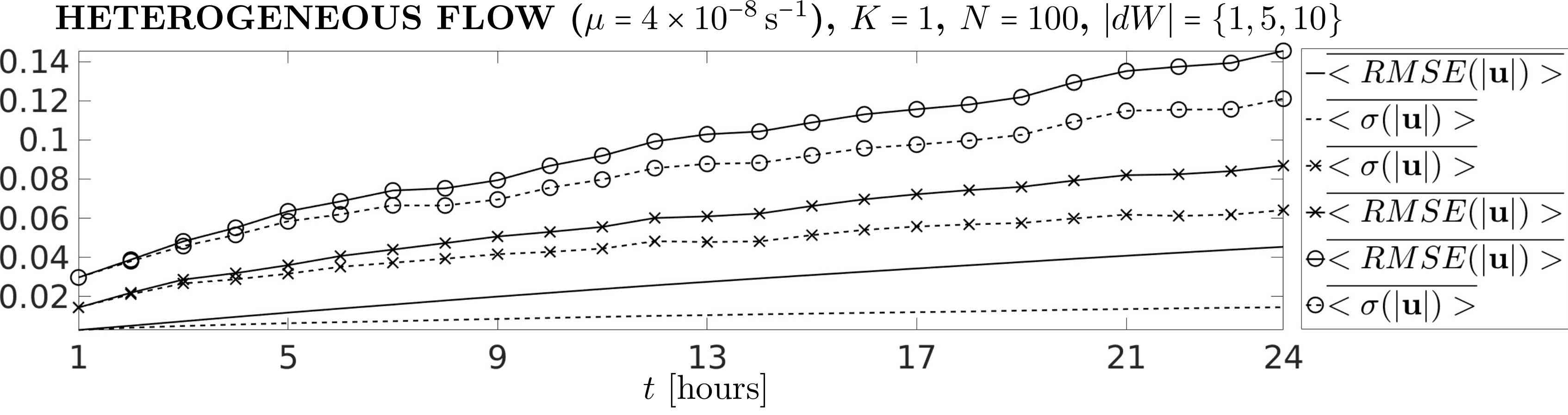}
\caption{Evolution of the root mean square error, $\overline{<RMSE(|\boldsymbol{u}|)>}$ (solid line),
and standard deviation, $\overline{<\sigma(|\boldsymbol{u}|)>}$ (dashed line), of the ensemble mean for 
the \textbf{\textit{heterogeneous flow}} computed for $K=1$,
$N=100$ and different amplitudes of the noise $dW$:  
$|dW|=1$ (solid/dashed line), $|dW|=5$ (solid/dashed line marked by a cross),
$|dW|=10$ (solid/dashed line marked by an circle).
\textit{Not that the initial ensemble is unbiased.}
\ansA{The results for the homogeneous flow look very similar.}}
\label{fig:rmse_std_xi1_2_4_100_200_400particles_mu4D-8_mu4D-7_new}
\end{center}
\end{figure}

Acting in the same vein, one can increase the amplitude of the noise which decreases the difference
between $\overline{<RMSE(|\boldsymbol{u}|)>}$ and $\overline{<\sigma(|\boldsymbol{u}|)>}$, 
but the ensemble still acquires a bias in time (Figure~\ref{fig:rmse_std_xi1_2_4_100_200_400particles_mu4D-8_mu4D-7_new}).

All these simulations with different parameters of the stochastic model show that the bias increases and eventually
becomes larger than the standard deviation. For example, for $T_{spin}=[-1,0]\, {\rm
  hour}$ and $K=1$, the bias remains smaller than the standard
deviation at 97\% of the spatial grid points 1 hour after the spin-up
time, but the proportion decreased to 34\% of the grid points after 20
time steps (10 hours).

A substantial improvement can be obtained by increasing the number of
driving Brownian motions. For example, if one uses
$K=\{32,64,128\}$, the bias remains smaller than the standard
deviation at 99\% of the spatial grid points 1 hour after the spin-up, but there is a substantial benefit after 20 time steps,
where the proportion of spatial grid points where bias remains less
than the standard deviation increases to $\{59\%,60\%,60.5\%\}$,
respectively. The proportion increases as we add more and more
Brownian motions up to a certain level. In this particular case,
$K=32$, it is close to the asymptotic value, which is approximately
60\%.

Another effect that we have investigated was the length of the spin-up
period. If instead of $T_{spin}=[-1,0]\, {\rm hour}$ we use, say,
$T_{spin}=[-12,0]\, {\rm hours}$ and $T_{spin}=[-24,0]\, {\rm hours}$,
and ($K=32$), the above proportions become 75\% and 81\%,
respectively.  All these experiments show that the quality of the
ensemble can be improved (when the "measure of goodness" is the
bias/standard deviation comparison) over a given period of time by
increasing both the number of Brownian motions driving the SPDE as
well as the spin-up time.
Another important conclusion is that
our choice of functions $\xi$ used in the stochastic parameterisation cannot deliver an unbiased solution.

\subsection{Uncertainty quantification for the deterministic QG model}

\begin{figure}[htp]
\begin{center}
\includegraphics[scale=0.135]{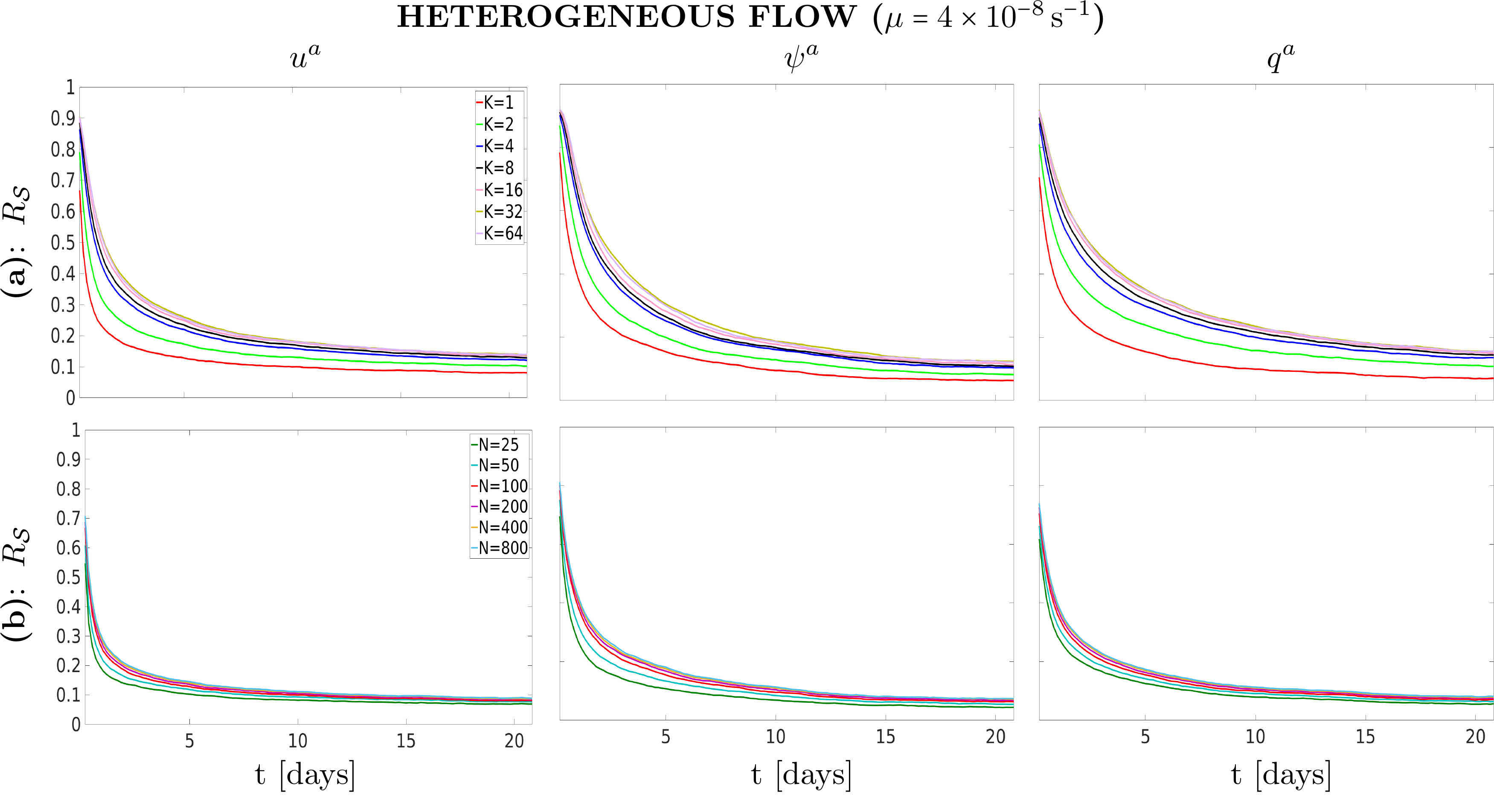}
\caption{Shown is the dependence of $R_{\mathcal{S}}$ (deterministic case)
for the velocity component $u^a$ ($v^a$ is not shown, since it behaves qualitatively similar to $u^a$), stream function $\psi^a$, and PV anomaly $q^a$ 
on {\bf (a)} the number of EOFs $K$ ($N=100$ in this case) 
and {\bf (b)} size of the stochastic ensemble $N$ ($K=1$ in this case) over the time period $T=[0,21]$ days for the 
\textbf{\textit{heterogeneous flow}} presented in Figure~\ref{fig:qf_qa_qc_qam_qcm_mu1D-8_129x65}; $G^c=129\times65$. 
The initial conditions for the deterministic model have been computed over the spin up period $T_{\rm spin}=[-8,0]$ hours.
\ansA{The results for the homogeneous flow look very similar.}
}
\label{fig:MC3test_mu4D-8_mu4D-7}
\end{center}
\end{figure}

Another important question we study is to quantify how the uncertainty in the initial stochastic condition is propagated
by the deterministic QG model and compare these uncertainty quantification results with the stochastic case \ansA{for the grids $G^c=\{129\times65$, $257\times129$\}}.
For doing so, we start the deterministic QG model from the stochastic initial condition at time $t=0$ (see Section~\ref{sec:ICs} for details)
and run it for each independent realization of the Brownian noise $W$, 
and compare the behavior of the deterministic (denoted by $R_{\mathcal{S}}$) 
and stochastic (denoted by $\widetilde{R}_{\mathcal{S}}$) spreads 
computed with the deterministic and stochastic QG models, respectively. The results of this simulation are given in Figure~\ref{fig:MC3test_mu4D-8_mu4D-7}.

\begin{figure}[htp]
\begin{center}
\includegraphics[scale=0.135]{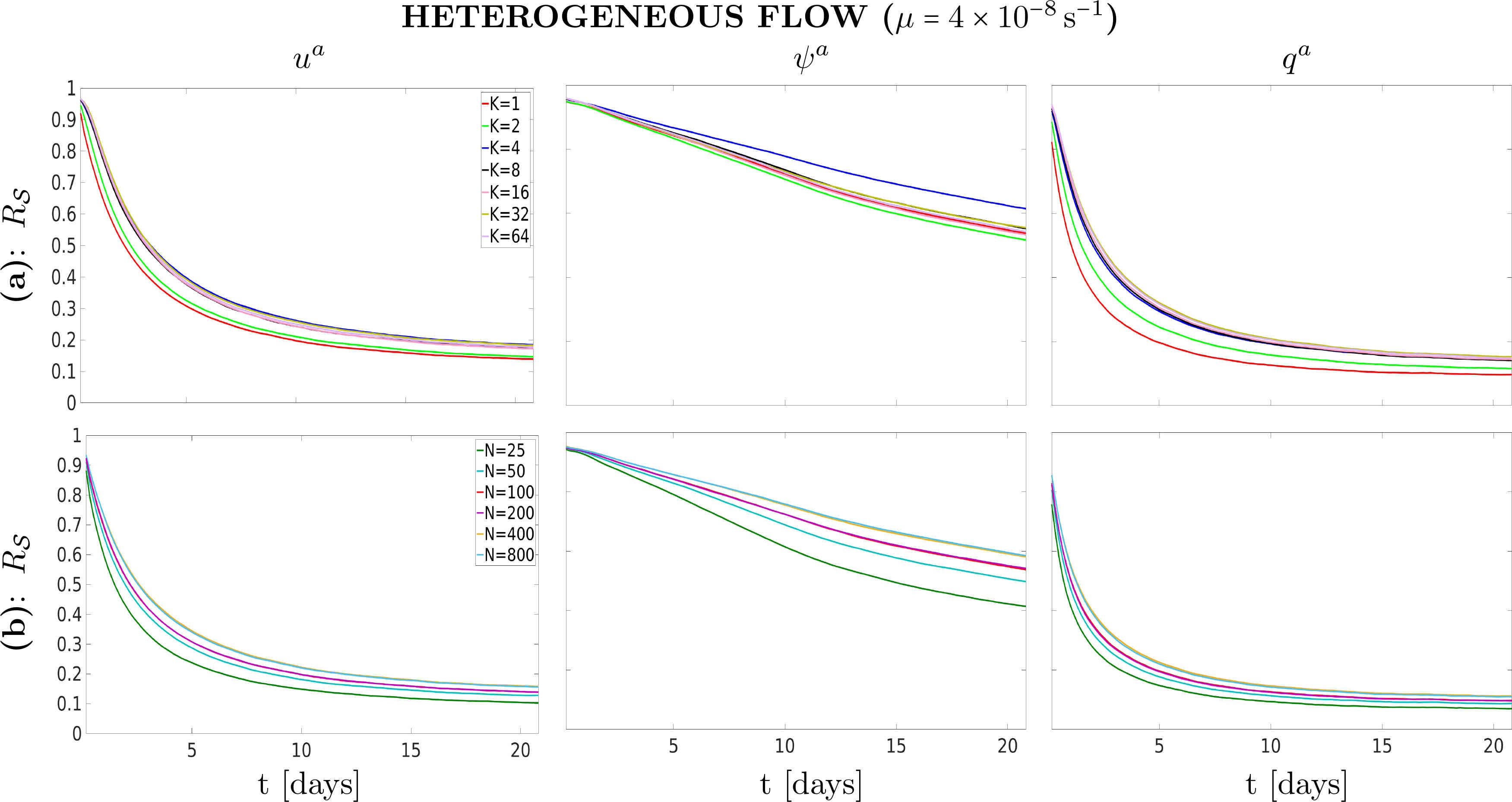}
\caption{The same as in Figure~\ref{fig:MC3test_mu4D-8_mu4D-7}, but for $G^c=257\times129$.
\ansA{The results for the homogeneous flow look very similar.}}
\label{fig:MC3test_mu4D-8_257x129}
\end{center}
\end{figure}

The results in Figure~\ref{fig:MC3test_mu4D-8_mu4D-7} clearly indicate that the ratio of points covered by the spread
in the deterministic case is much less compared to the stochastic case
(compare Figures~\ref{fig:MC3test_mu4D-8_mu4D-7} and~\ref{fig:R_EOF_ensemble_mu4D-8}). 
For the higher resolution ($G^c=257\times129$), the situation is similar (Figure~\ref{fig:MC3test_mu4D-8_257x129}), although more
points in the domain remain within the spread compared with the lower-resolution simulation.
These simulations show that the spread of stochastic solutions can track the true solution for a longer time 
compared with the deterministic model, thus providing solid foundations for data assimilation methods, which require the "truth" to be contained within the spread of 
stochastic solutions unless nudging methods are used.

\begin{figure}[htp]
\begin{center}
\includegraphics[scale=0.135]{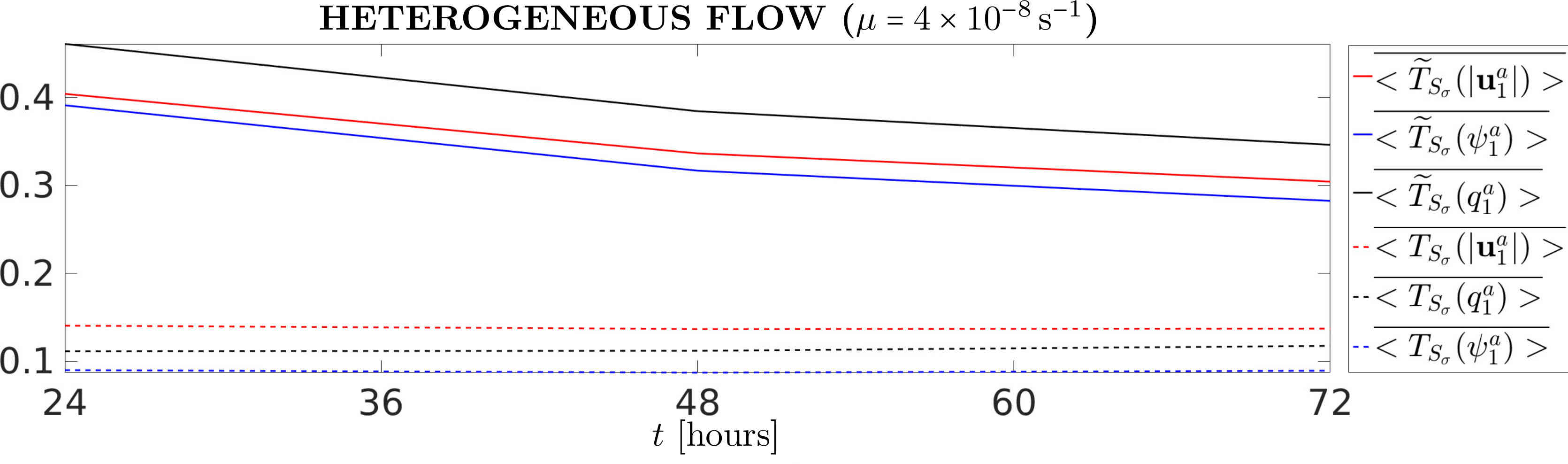}
\caption{Evolution of $\overline{<\widetilde{T}_{\mathcal{S}_{\sigma}}>}$ (solid line; stochastic case) and 
$\overline{<T_{\mathcal{S}_{\sigma}}>}$ (dashed line; deterministic case) for 
the heterogeneous flow; the number of leading EOFs is $K=1$, and the size of the ensemble
is $N=100$.
\ansA{The results for the homogeneous flow look very similar.}}
\label{fig:std_TimeCoveredBySpread_xi1_100particles_mu4D-8_mu4D-7__stochastic_vs_deterministic}
\end{center}
\end{figure}

\ansA{To examine the length of time for which}
the true deterministic solution lie within one standard deviation of the stochastic 
(measured by $\overline{<\widetilde{T}_{\mathcal{S}_{\sigma}}>}$) 
and deterministic (measured by $\overline{<T_{\mathcal{S}_{\sigma}}>}$) \ansA{ensembles.
Figure~\ref{fig:std_TimeCoveredBySpread_xi1_100particles_mu4D-8_mu4D-7__stochastic_vs_deterministic} shows the evolution 
of these quantities in time.}

As shown in Figure~\ref{fig:std_TimeCoveredBySpread_xi1_100particles_mu4D-8_mu4D-7__stochastic_vs_deterministic},
the spread computed with the stochastic model keeps track of the true solution over a longer period
of time compared to the one computed with the deterministic model
(compare the solid lines (which correspond to $\widetilde{T}_{\mathcal{S}_{\sigma}}$; stochastic case) 
with the dashed lines (which correspond to $T_{\mathcal{S}_{\sigma}}$; deterministic case) 
in Figure~\ref{fig:std_TimeCoveredBySpread_xi1_100particles_mu4D-8_mu4D-7__stochastic_vs_deterministic}).
This conclusion holds true for both the homogeneous and heterogeneous flows. 
It is important to note that
despite the fact that the stochastic model has been calibrated for 24 hours (section~\ref{sec:approx_lagrangian_evol}), 
it gives very good results well beyond this time period for both heterogeneous and homogeneous flows.

\begin{figure}
\centering
\hspace*{-0.5cm}
\includegraphics[scale=0.25]{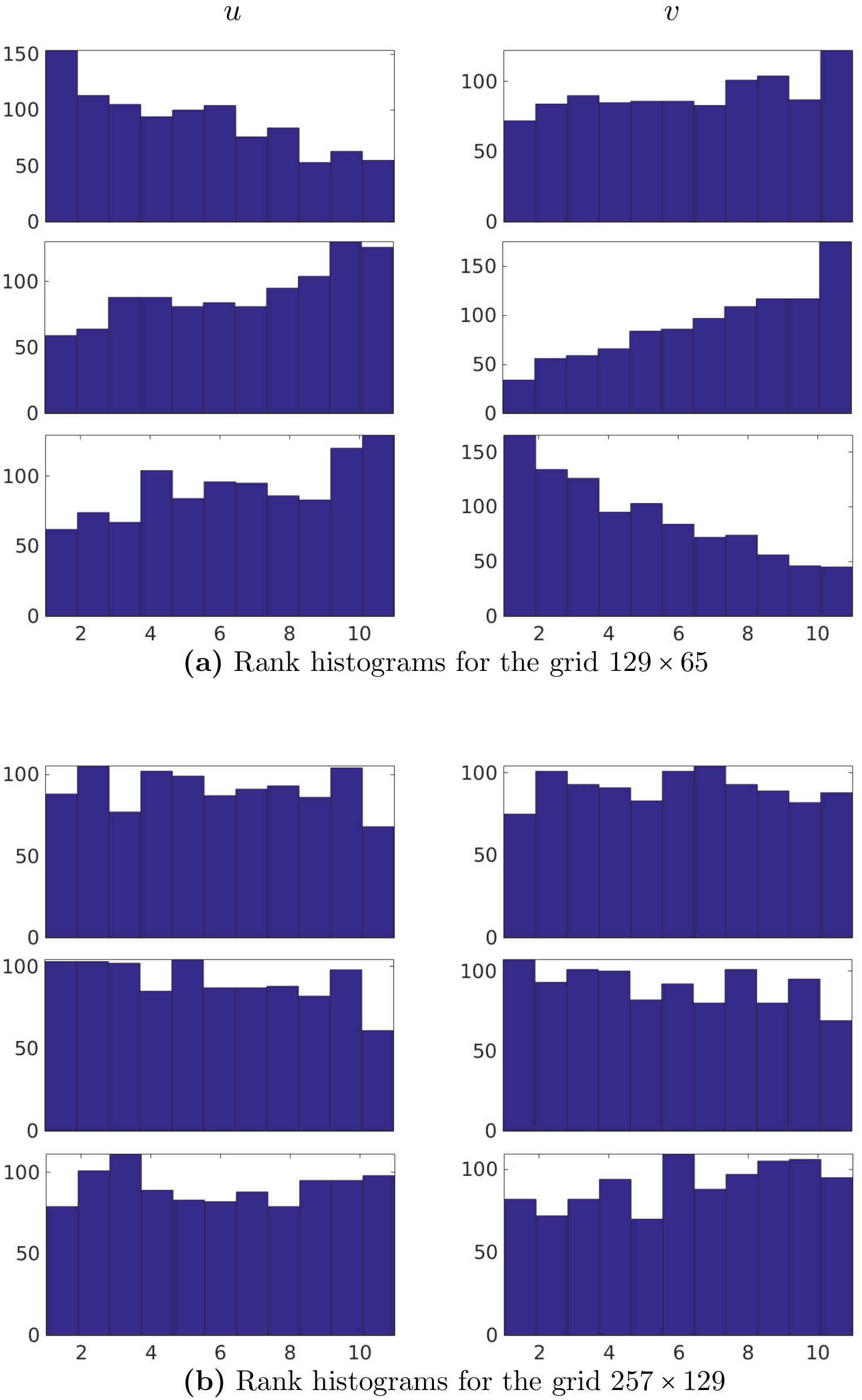}
\caption{\ansBB{Typical rank histograms for heterogeneous flow velocity $\mathbf{u}=(u,v)$ at different locations (not shown) and resolutions.
Note that a pair of histograms at different resolutions shares the same location in the computation domain.
Each histogram is based on 1000 forecast-observation pairs generated by solving the stochastic QG model. 
For simulating the stochastic QG model we use 100 stochastic solutions and 32 leading EOFs.
Each stochastic solution for the rank histogram is selected randomly from the ensemble every 4 hours. The rank histograms at the higher resolution are flatter and t hus the forecasts are more reliable, although this resolution is still much less than the reference simulation.}}
\label{fig:rank_histograms}
\end{figure}

\ansBB{In addition, we have studied how the space resolution affects the ensemble reliability 
by analysing forecast reliability rank histograms (Talagrand diagrams) for the stochastic QG model at different resolutions (Figure~\ref{fig:rank_histograms}).
In particular, we have applied the approach used in~\cite{Brocker2018}. Namely, the rank histograms are computed by randomly selecting 10 stochastic solutions out of the 
total of 100 every 4 hours and comparing the new observations with the stochastic solution perturbed by the noise with the same amplitude as the measurement noise.}       

\ansBB{As can be seen in Figure~\ref{fig:rank_histograms}, the stochastic ensemble computed on the grid $129\times65$ has a bias
(rank histograms are not flat; see Figure~\ref{fig:rank_histograms}a) at all observation locations thus making the ensemble prediction unreliable. 
However, the situation changes for the better at the slightly less coarse $257\times129$ (Figure~\ref{fig:rank_histograms}b).
In particular, the rank histograms becomes flatter and are thus less biased. In order to correct the bias which is still there even at the higher resolution,
and to produce a reliable forecast, we have proposed a data assimilation methodology which we investigated in the sequel paper~\cite{CCHWS2020_1}.
}

\section{Deterministic vs Stochastic\label{sec:deter_vs_stoch}}
\begin{figure}[htp]
\begin{center}
\includegraphics[scale=0.14]{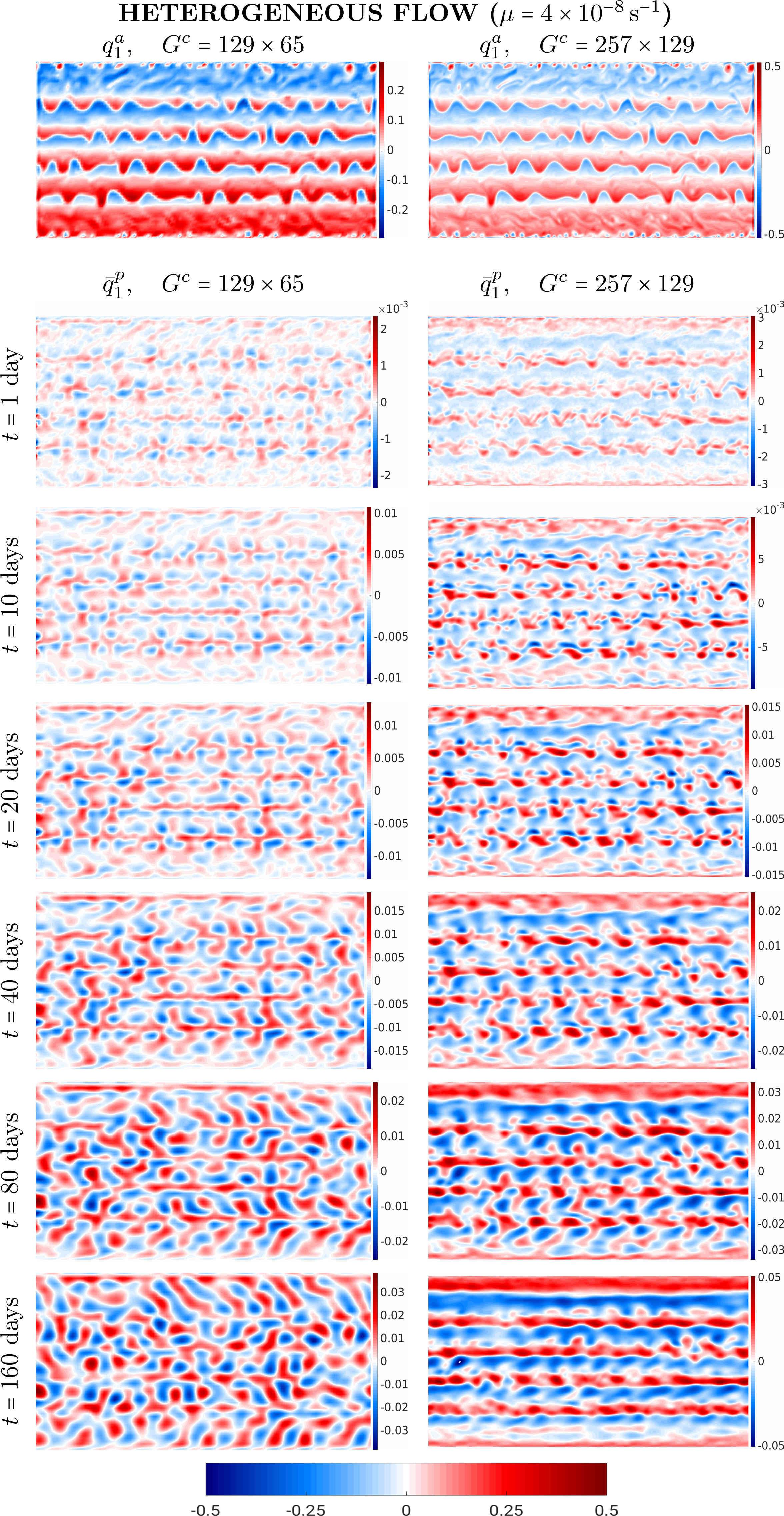}
\caption{The series of snapshots shows the true deterministic solution $q^a_1$ 
computed on $G^c=\{129\times65$, $257\times129$\}, and 
the low-resolution parameterised solution $\bar{q}^p_1$ (averaged over the stochastic ensemble of size $N=100$)
computed on the same coarse grids 
by simulating the stochastic QG model from a randomly perturbed zero initial condition
for the \textbf{\textit{heterogeneous flow}};
the parameterised solution uses 32 leading EOFs.
All solutions are given in units of $[s^{-1}f^{-1}_0]$,  where $f_0=0.83\times10^{-4}\, {\rm s^{-1}}$ is the Coriolis parameter.
The true solutions correspond to day 1 and presented here just to show the structure of the flow.
}
\label{fig:deter_vs_stoch_mu4D-8}
\end{center}
\end{figure}

In this section we compare the deterministic QG equation with its stochastic parameterisation with the goal to analyse how well 
the stochastic model works. From now on, we only focus on the heterogeneous flow.
It is instructive to firstly analyse whether the parameterisation can restore the number of jets
when the model starts from a randomly perturbed zero initial condition (Figure~\ref{fig:deter_vs_stoch_mu4D-8}).

From Figure~\ref{fig:deter_vs_stoch_mu4D-8} we conclude that at the low resolution ($G^c=129\times65$)
the parameterisation cannot restore the correct number of jets, and therefore can lead to poor representation of
the effects of the small scale dynamics on the resolved scales. 
We hypothesize that this can be due to the fact that the very low energy of the stochastic solution is not enough to properly
shape the jets. In particular, the time averaged ensemble mean total energy is $\bar{\left<E^{129\times65}\right>}=0.01$.
In contrast, at the higher resolution ($G^c=257\times129$)
the parameterisation demonstrates a much better performance, and restores the original number of jets. 
In this case, the energy is two orders of magnitude higher compared to the lower resolution case, 
$\bar{\left<E\right>}^{257\times129}=3.9$.
%

\begin{figure}[htp]
\begin{center}
\includegraphics[scale=0.135]{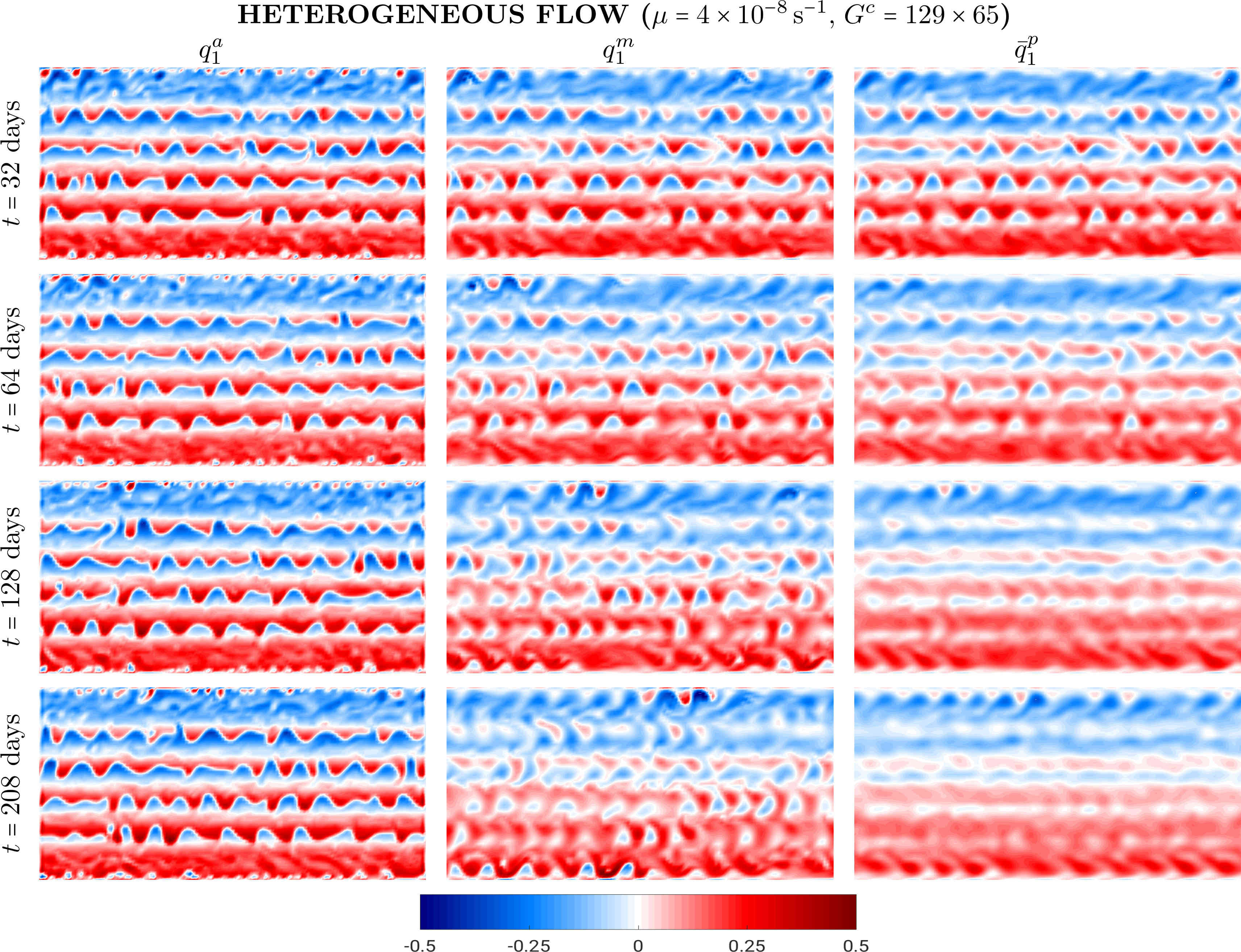}
\caption{The series of snapshots shows the true deterministic solution $q^a_1$, modelled solution $q^m_1$ 
computed with the deterministic QG model,
and parameterised solution $\bar{q}^p_1$ (averaged over the stochastic ensemble of size $N=100$) computed with the stochastic
QG model. All solutions were computed on the same grid $G^c=129\times65$, have the same initial condition,
and the parameterised solution uses 32 leading EOFs.
All fields are given in units of $[s^{-1}f^{-1}_0]$,  where $f_0=0.83\times10^{-4}\, {\rm s^{-1}}$ is the Coriolis parameter.
}
\label{fig:deter_vs_stoch_mu4D-8_129x65}
\end{center}
\end{figure}

\begin{figure}[htp]
\begin{center}
\includegraphics[scale=0.135]{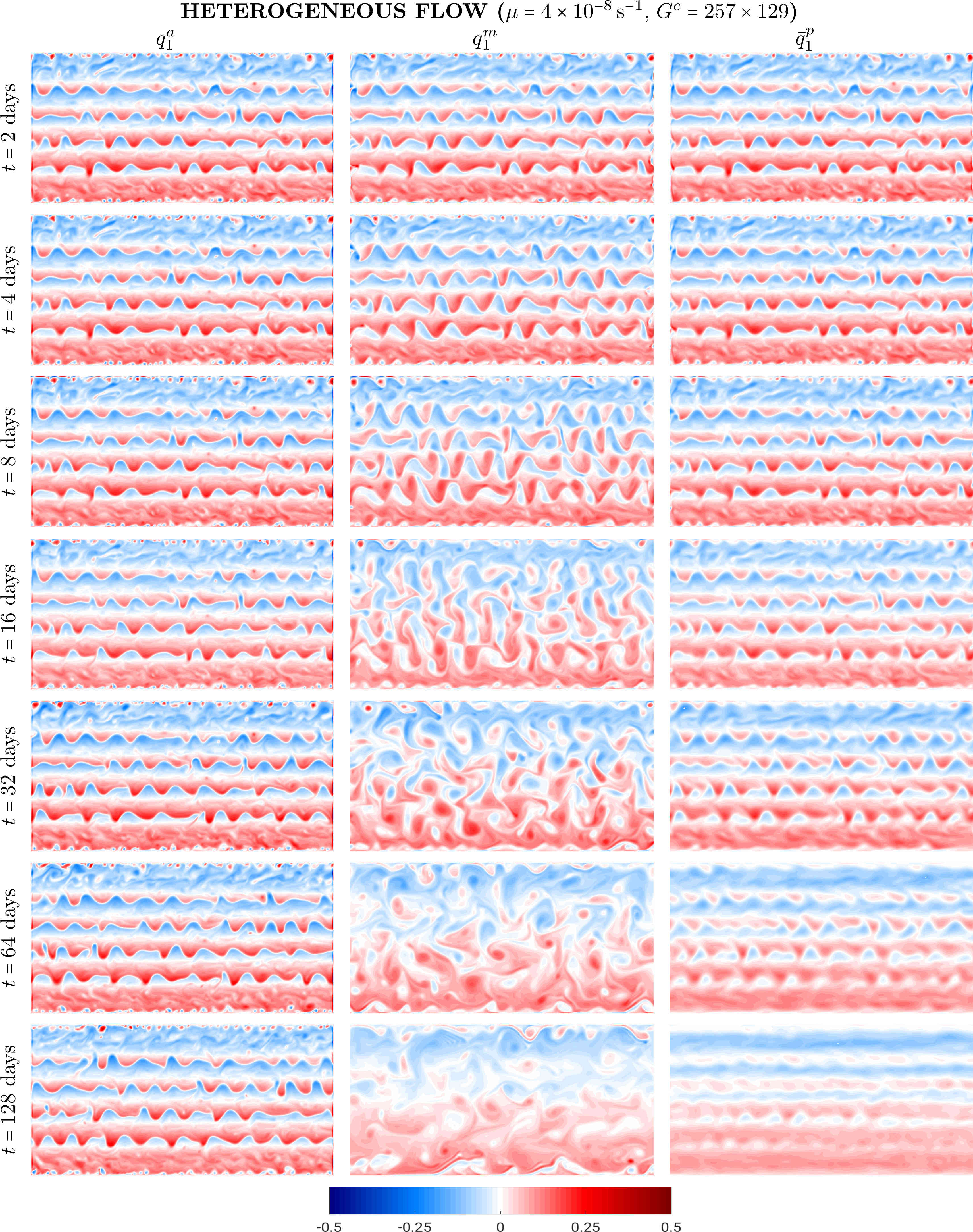}
\caption{
The same as in Figure~\ref{fig:deter_vs_stoch_mu4D-8_129x65}, but for $G^c=257\times129$.
}
\label{fig:deter_vs_stoch_mu4D-8_257x129}
\end{center}
\end{figure}

In order to estimate the efficacy of the parameterisation we study how it performs at different resolutions
and how accurate it can reproduce the truth. The results are presented in Figures~\ref{fig:deter_vs_stoch_mu4D-8_129x65}
and~\ref{fig:deter_vs_stoch_mu4D-8_257x129}.

The results in Figure~\ref{fig:deter_vs_stoch_mu4D-8_129x65} are shown for day 32 onwards, since all three solutions
$q^a_1$, $q^m_1$, and $\bar{q}^p_1$ look very similar before that day. As seen in Figure~\ref{fig:deter_vs_stoch_mu4D-8_129x65},
both the deterministic QG model (middle column) and 
its stochastic parameterisation (right column) gradually diffuse 
out the jets and ruin the structure of the flow. However, when the resolution is higher ($G^c=257\times129$) the situation
is different (Figure~\ref{fig:deter_vs_stoch_mu4D-8_257x129}). In particular, the 
coarse-resolution deterministic model cannot maintain the jet-like structure of the flow and tries to switch to the
two-jet regime it originally had (middle column). 
In contrast, the stochastic QG model can maintain the original structure of the flow over a much longer period of time
(right column) compared with the lower resolution model.

\begin{figure}[htp]
\begin{center}
\includegraphics[scale=0.135]{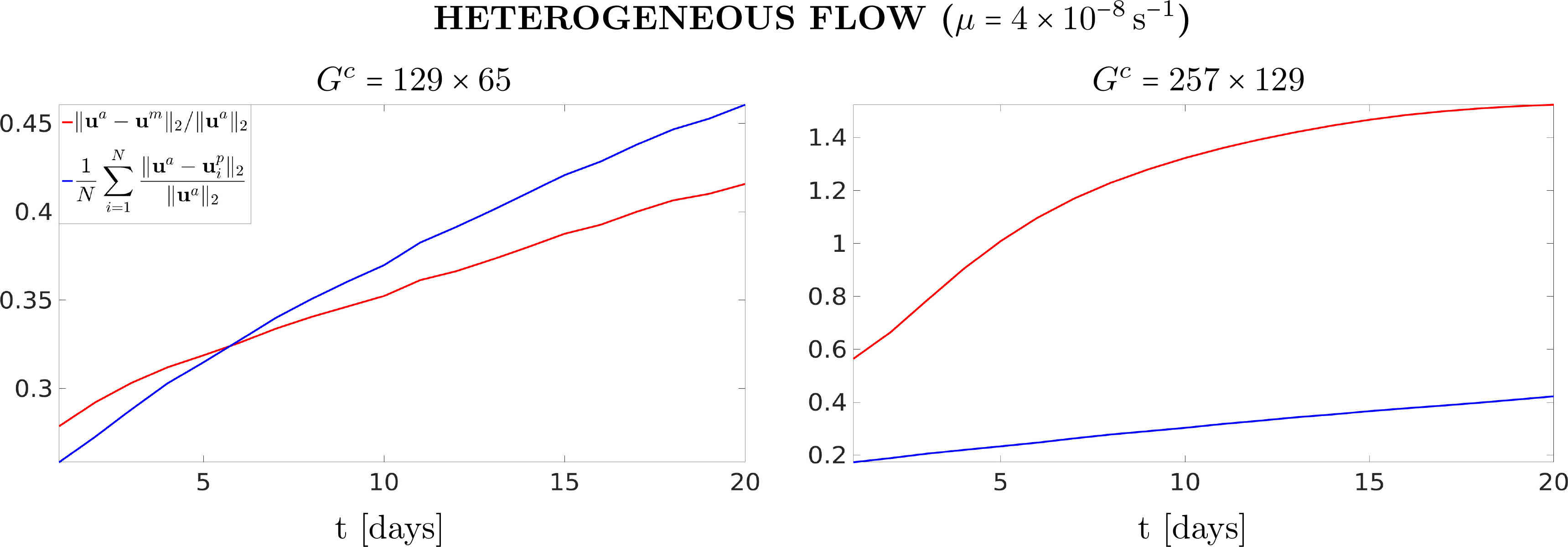}
\caption{Shown is the relative $l_2$-norm error between the true deterministic solution, ${\bf u}^a$, and 
modelled solution ${\bf u}^m$ (red line) and parameterised solution ${\bf u}^p$ (blue line) as a function of time.
}
\label{fig:deter_vs_stoch_mu4D-8_129x65_257x129}
\end{center}
\end{figure}

Analysis of the relative $l_2$-norm error between the true solution and its approximations (modelled solution ${\bf u}^m$ and 
parameterised solution ${\bf u}^p$)
shows that the deterministic model on the grid $G^c=129\times65$ outperforms the stochastic parameterisation
(Figure~\ref{fig:deter_vs_stoch_mu4D-8_129x65_257x129}, left). In particular, 
after approximately 7 days the parameterised solution (blue line) starts loosing the track of the truth, while the deterministic QG
model started from the same stochastic initial conditions gives better results (red line).
When the resolution is higher ($G^c=257\times129$) the parameterised model performs much better than the deterministic one
(Figure~\ref{fig:deter_vs_stoch_mu4D-8_129x65_257x129}, right). We hypothesise that at higher resolutions small scales
are better energized thus intensifying the backscatter mechanism (e.g,~\cite{SB2016}) which, in turn, channels the energy into the zonal jets.

In summary, by comparing the uncertainty quantification results for the heterogeneous and homogeneous flow, 
we have found that the stochastic spread captures 
the true deterministic solution $q^a$ (computed on the coarse grid $G^c$) 
for much more individual grid points in the computational domain (Figure~\ref{fig:MC3test_mu4D-8_mu4D-7})
and for much longer times (Figure~\ref{fig:std_TimeCoveredBySpread_xi1_100particles_mu4D-8_mu4D-7__stochastic_vs_deterministic}) 
than the deterministic spread 
(compare either the stream function $\psi^a_1$, velocity $\boldsymbol{u}^a_1$, or 
PV anomaly $q^a_1$ for the stochastic spread (top row) and deterministic spread (bottom row) 
in Figures~\ref{fig:MC3test_mu4D-8_mu4D-7} and~\ref{fig:std_TimeCoveredBySpread_xi1_100particles_mu4D-8_mu4D-7__stochastic_vs_deterministic}).
Therefore,  for data assimilation, the proposed parameterisation would be considerably more preferable than the deterministic QG model; 
not only for heterogeneous flows, but also for homogeneous ones. 
This study has also shown that the higher the resolution is, the better the stochastic parameterisation performs.
Overall, we conclude that the parameterisation of the stochastic QG model is robust to large 
variations of the flow dynamics and governing parameters, and can be equally well applied to both homogeneous and heterogeneous flows.

\section{Conclusion and future work}
\label{sec:concl}
In this paper we have introduced a stochastic parameterisation for
unresolved eddy motions in a two-layer quasi-geostrophic channel model
with forcing and dissipation. The parameterisation is based upon
the idea of ``transport noise'', which models the modifications to the
velocity field due to unresolved dynamics. This model assumes that the
transport of large scale components is accurate, but that the velocity
field used to transport these components is missing contributions from
unresolved scales.
We have proposed a method for extracting stochastic forcing by post-processing
high resolution simulations, and demonstrated it by using uncertainty
quantification experiments for both the SDE and stochastic QG model for homogeneous and heterogeneous
flow dynamics. We have also shown that the quality of the stochastic ensemble 
can be improved by increasing both the number of Brownian motions driving the SPDE as
well as the spin-up time. However, 
our choice of functions $\xi$ used in the stochastic parameterisation cannot deliver an unbiased solution, and therefore
it requires a bias correction, which can be done using data assimilation methods. Another important finding is that higher 
resolutions significantly improve the parameterised solution.
Overall, our results show that the proposed parameterisation
significantly depends on the resolution of the stochastic model and 
gives good ensemble performance for both
homogeneous and heterogeneous flows, and they lay the solid foundation for data assimilation.

In future work, we intend to use this approach as the basis for data
assimilation algorithms, to investigate the assimilation of data from
a high-resolution deterministic model into a low-resolution stochastic
model. We also intend \ansA{to examine the derivation of} the ``prognostic''
parameterisations derived in \cite{GBHo2018} in which the stochastic forcing patterns are determined
from the coarse model itself using physical principles, rather than
the ``diagnostic'' parameterisations of this paper where they are
determined from high resolution simulations. The diagnostic approach
proposed in this paper will provide important insight by comparing the
diagnosed forcing with the state of the stochastic model.

\section*{Acknowledgments}
The authors thank The Engineering and Physical Sciences Research Council for the support of this work through the grant EP/N023781/1.
The work of the second author has also been partially supported by a UC3M-Santander Chair of Excellence grant held at the Universidad Carlos III de Madrid.
We thank Pavel Berloff, Mike Cullen, John Gibbon, Georg Gottwald, Nikolas Kantas, Etienne Memin, 
Sebastian Reich, Valentin Resseguier, and Aretha Teckentrup for useful and valuable discussions throughout the preparation of this work. 
We thank the anonymous referees for their constructive comments and suggestions, which have helped us to improve the paper.

\bibliographystyle{apalike}

\begin{thebibliography}{99}

\bibitem[Abramov, 2016]{Abramov2016}
Abramov, R. (2016).
\newblock A simple stochastic parameterization for reduced models of multiscale
  dynamics.
\newblock {\em Fluids}, 1:1--18.

\bibitem[Arakawa, 1966]{Arakawa1966}
Arakawa, A. (1966).
\newblock Computational design for long-term numerical integration of the
  equations of fluid motion: two-dimensional incompressible flow. {Part I}.
\newblock {\em J. Comput. Phys.}, 1:119--143.

\bibitem[Berloff, 2005]{Berloff2005}
Berloff, P. (2005).
\newblock Random-forcing model of the mesoscale oceanic eddies.
\newblock {\em J. Fluid Mech.}, 529:71--95.

\bibitem[Berloff and Kamenkovich, 2013]{BerloffKamenkovich2013}
Berloff, P. and Kamenkovich, I. (2013).
\newblock On spectral analysis of mesoscale eddies. {P}art {I}: {L}inear
  analysis.
\newblock {\em J. Phys. Oceanogr.}, 43:2505--2527.

\bibitem[Berloff et~al., 2011]{Berloff_et_al2011}
Berloff, P., Karabasov, S., Farrar, T., and Kamenkovich, I. (2011).
\newblock On latency of multiple zonal jets in the oceans.
\newblock {\em J. Fluid Mech.}, 686:534--567.

\bibitem[Berloff and McWilliams, 2002]{BM2002}
Berloff, P. and McWilliams, J. (2002).
\newblock {Material transport in oceanic gyres. Part II: Hierarchy of
  stochastic models.}
\newblock {\em J. Phys. Oceanogr.}, 32:797--830.

\bibitem[Berloff and McWilliams, 2003]{BM2003}
Berloff, P. and McWilliams, J. (2003).
\newblock {Material transport in oceanic gyres. Part III: Randomized stochastic
  models.}
\newblock {\em J. Phys. Oceanogr.}, 33:1416--1445.

\bibitem[Br\"ocker, 2018]{Brocker2018}
Br\"ocker, J. (2018).
\newblock Assessing the reliability of ensemble forecasting systems under
  serial dependence.
\newblock {\em Q. J. R. Meteorol. Soc}, 144:2666--2675.

\bibitem[Chen et~al., 2012]{Chen_et_al2012}
Chen, K., Tu, J., and Rowley, C. (2012).
\newblock Variants of dynamic mode decomposition: connections between {K}oopman
  and {F}ourier analyses.
\newblock {\em J. Nonlinear Sci.}, 22:887--915.

\bibitem[Cooper and Zanna, 2015]{CooperZanna2015}
Cooper, F. and Zanna, L. (2015).
\newblock Optimization of an idealised ocean model, stochastic parameterisation
  of sub-grid eddies.
\newblock {\em Ocean Model.}, 88:38--53.

\bibitem[Cotter et~al., 2020]{CCHWS2020_1}
Cotter, C., Crisan, D., Holm, D., Pan, W., and Shevchenko, I. (2020).
\newblock Data assimilation for a quasi-geostrophic model with
  circulation-preserving stochastic transport noise.
\newblock {\em J. Stat. Phys.}
\newblock https://doi.org/10.1007/s10955-020-02524-0.

\bibitem[Cotter et~al., 2017]{CoGoHo2017}
Cotter, C., Gottwald, G., and Holm, D. (2017).
\newblock Stochastic partial differential fluid equations as a diffusive limit
  of deterministic lagrangian multi-time dynamics.
\newblock {\em Proc. Roy. Soc. A}, 473.

\bibitem[Duan and Nadiga, 2007]{DuanNadiga2007}
Duan, J. and Nadiga, B. (2007).
\newblock Stochastic parameterization for large eddy simulation of geophysical
  flows.
\newblock {\em Proc. Am. Math. Soc.}, 135:1187--1196.

\bibitem[Elsner and Tsonis, 1996]{ElsnerTsonis1996}
Elsner, J. and Tsonis, A. (1996).
\newblock {\em {Singular Spectrum Analysis: A New Tool in Time Series
  Analysis}}.
\newblock Plenum Press, New York.

\bibitem[Franzke et~al., 2005]{franzke2005low}
Franzke, C., Majda, A.~J., and Vanden-Eijnden, E. (2005).
\newblock Low-order stochastic mode reduction for a realistic barotropic model
  climate.
\newblock {\em Journal of the atmospheric sciences}, 62(6):1722--1745.

\bibitem[Frederiksen et~al., 2012]{Frederiksen_et_al2012}
Frederiksen, J., O’Kane, T., and Zidikheri, M. (2012).
\newblock Stochastic subgrid parameterizations for atmospheric and oceanic
  flows.
\newblock {\em Phys. Scr.}, 85:068202.

\bibitem[Gay-Balmaz and Holm, 2018]{GBHo2018}
Gay-Balmaz, F. and Holm, D.~D. (2018).
\newblock Stochastic geometric models with non-stationary spatial correlations
  in lagrangian fluid flows.
\newblock {\em J. Nonlinear Sci.}, 28:873--904.

\bibitem[Gent, 2011]{Gent2011}
Gent, P. (2011).
\newblock The {Gent–McWilliams} parameterization: 20/20 hindsight.
\newblock {\em Ocean Model.}, 39:2--9.

\bibitem[Gent and Mcwilliams, 1990]{GentMcwilliams1990}
Gent, P. and Mcwilliams, J. (1990).
\newblock Isopycnal mixing in ocean circulation models.
\newblock {\em J. Phys. Oceanogr.}, 20:150--155.

\bibitem[Grooms et~al., 2015]{Grooms_et_al2015}
Grooms, I., Majda, A., and Smith, K. (2015).
\newblock Stochastic superparametrization in a quasigeostrophic model of the
  {Antarctic Circumpolar Current}.
\newblock {\em Ocean Model.}, 85:1--15.

\bibitem[Hairer et~al., 1993]{HNW1993}
Hairer, E., N\o{}rsett, S., and Wanner, G. (1993).
\newblock {\em Solving Ordinary Differential Equations {I}. {Nonstiff
  Problems}}.
\newblock Springer, Berlin, Heidelberg.

\bibitem[Hannachi et~al., 2007]{HaJoSt2007}
Hannachi, A., Jolliffe, I., and Stephenson, D. (2007).
\newblock Empirical orthogonal functions and related techniques in atmospheric
  science: {A} review.
\newblock {\em Int. J. Climatol.}, 27:1119--1152.

\bibitem[Holm, 2015]{holm2015variational}
Holm, D. (2015).
\newblock Variational principles for stochastic fluids.
\newblock {\em Proc. Roy. Soc. A}, 471.

\bibitem[Hundsdorfer et~al., 1995]{Hundsdorfer_et_al1995}
Hundsdorfer, W., Koren, B., van Loon, M., and Verwer, J. (1995).
\newblock A positive finite-difference advection scheme.
\newblock {\em J. Comput. Phys.}, 117:35--46.

\bibitem[Kamenkovich et~al., 2009]{Kamenkovich_et_al2009}
Kamenkovich, I., Berloff, P., and Pedlosky, J. (2009).
\newblock Anisotropic material transport by eddies and eddy-driven currents in
  a model of the {N}orth {A}tlantic.
\newblock {\em J. Phys. Oceanogr.}, 39:3162--3175.

\bibitem[Karabasov et~al., 2009]{Karabasov_et_al2009}
Karabasov, S., Berloff, P., and Goloviznin, V. (2009).
\newblock {CABARET} in the ocean gyres.
\newblock {\em Ocean Model.}, 2--3:155--168.

\bibitem[Karatzas and Shreve, 1991]{KaratzasShreve1991}
Karatzas, I. and Shreve, S. (1991).
\newblock {\em Brownian motion and stochastic calculus}.
\newblock Graduate Texts in Mathematics, 113. Springer-Verlag, New York.

\bibitem[Kraichnan, 1968]{Kraichnan1968}
Kraichnan, R. (1968).
\newblock Small-scale structure of a randomly advected passive scalar.
\newblock {\em Phys. Rev. Lett.}, 11:945--963.

\bibitem[Leith, 1990]{Leith1990}
Leith, C. (1990).
\newblock Stochastic backscatter in a subgrid‐scale model: Plane shear mixing
  layer.
\newblock {\em Phys. Fluids A: Fluid Dynamics}, 2:297--299.

\bibitem[Leutbecher et~al., 2017]{Leutbecher_et_al2017}
Leutbecher, M., Lock, S.-J., Ollinaho, P., Lang, S., Balsamo, G., Bechtold, P.,
  Bonavita, M., Christensen, H., Diamantakis, M., Dutra, E., English, S.,
  Fisher, M., Forbes, R., Goddard, J., Haiden, T., Hogan, R., Juricke, S.,
  Lawrence, H., MacLeod, D., Magnusson, L., Malardel, S., Massart, S., Sandu,
  I., Smolarkiewicz, P., Subramanian, A., Vitart, F., Wedi, N., and Weisheimer,
  A. (2017).
\newblock Stochastic representations of model uncertainties at {ECMWF}: state
  of the art and future vision.
\newblock {\em Q. J. R. Meteorol. Soc.}, 143:2315--2339.

\bibitem[Majda et~al., 2001]{Majda_et_al2001}
Majda, A., Timofeyev, I., and Vanden-Eijnden, E. (2001).
\newblock A mathematical framework for stochastic climate models.
\newblock {\em Comm. Pure Appl. Math.}, 54:891--974.

\bibitem[Mana and Zanna, 2014]{PortaMana_Zanna2014}
Mana, P. and Zanna, L. (2014).
\newblock Toward a stochastic parameterization of ocean mesoscale eddies.
\newblock {\em Ocean Model.}, 79:1--20.

\bibitem[McWilliams, 1977]{McWilliams1977}
McWilliams, J. (1977).
\newblock A note on a consistent quasigeostrophic model in a multiply connected
  domain.
\newblock {\em Dynam. Atmos. Ocean}, 5:427--441.

\bibitem[M{\'e}min, 2014]{Memin2014}
M{\'e}min, E. (2014).
\newblock Fluid flow dynamics under location uncertainty.
\newblock {\em Geophys. Astrophys. Fluid Dyn.}, 108:119--146.

\bibitem[Pedlosky, 1987]{Pedlosky1987}
Pedlosky, J. (1987).
\newblock {\em Geophysical fluid dynamics}.
\newblock Springer-Verlag, New York.

\bibitem[Preisendorfer, 1988]{Preisendorfer1988}
Preisendorfer, R.~W. (1988).
\newblock {\em Principal Component Analysis in Meteorology and Oceanography}.
\newblock Elsevier, Amsterdam.

\bibitem[Schmid, 2010]{Schmid2010}
Schmid, P. (2010).
\newblock Dynamic mode decomposition of numerical and experimental data.
\newblock {\em J. Fluid Mech.}, 656:5--28.

\bibitem[Shevchenko and Berloff, 2015]{SB2015}
Shevchenko, I. and Berloff, P. (2015).
\newblock Multi-layer quasi-geostrophic ocean dynamics in eddy-resolving
  regimes.
\newblock {\em Ocean Modell.}, 94:1--14.

\bibitem[Shevchenko and Berloff, 2016]{SB2016}
Shevchenko, I. and Berloff, P. (2016).
\newblock Eddy backscatter and counter-rotating gyre anomalies of midlatitude
  ocean dynamics.
\newblock {\em Fluids}, 1:1--16.

\bibitem[Shu and Osher, 1988]{ShuOsher1988}
Shu, C. and Osher, S. (1988).
\newblock Efficient implementation of essentially non-oscillatory
  shock-capturing schemes.
\newblock {\em J. Comput. Phys.}, 77:439--471.

\bibitem[Siegel et~al., 2001]{SiegelEtAl2001}
Siegel, A., Weiss, J., Toomre, J., McWilliams, J., Berloff, P., and Yavneh, I.
  (2001).
\newblock Eddies and vortices in ocean basin dynamics.
\newblock {\em Geophys. Res. Lett.}, 28:3183--3186.

\bibitem[Vallis, 2006]{Vallis2006}
Vallis, G. (2006).
\newblock {\em Atmospheric and oceanic fluid dynamics: {F}undamentals and
  large-scale circulation}.
\newblock Cambridge University Press, Cambridge, UK.

\bibitem[Vannitsem, 2014]{vannitsem2014stochastic}
Vannitsem, S. (2014).
\newblock Stochastic modelling and predictability: analysis of a low-order
  coupled ocean--atmosphere model.
\newblock {\em Philosophical Transactions of the Royal Society A: Mathematical,
  Physical and Engineering Sciences}, 372(2018):20130282.

\bibitem[Weigel, 2012]{Weigel2012}
Weigel, A. (2012).
\newblock Ensemble forecasts.
\newblock In Jolliffe, I. and Stephenson, D., editors, {\em Forecast
  Verification: A Practitioner's Guide in Atmospheric Science}, chapter~8,
  pages 141--166. John Wiley \& Sons, Oxford, UK, 2 edition.

\bibitem[Woodward and Colella, 1984]{WoodwardColella1984}
Woodward, P. and Colella, P. (1984).
\newblock The numerical simulation of two-dimensional fluid flow with strong
  shocks.
\newblock {\em J. Comput. Phys.}, 54:115--173.

\bibitem[Wouters and Lucarini, 2012]{wouters2012disentangling}
Wouters, J. and Lucarini, V. (2012).
\newblock Disentangling multi-level systems: averaging, correlations and
  memory.
\newblock {\em Journal of Statistical Mechanics: Theory and Experiment},
  2012(03):P03003.

\end{thebibliography}

\medskip
Received xxxx 20xx; revised xxxx 20xx.
\medskip

\end{document}